\documentclass[twocolumn,aps,showpacs,prb,usenames,superscriptaddress]{revtex4-1}

\usepackage{float}
\restylefloat{table}
\usepackage{adjustbox}
\usepackage{tikz}
\usetikzlibrary{calc,positioning}
\usepackage{amsmath,amssymb}
\usepackage{subfig}
\usepackage{graphicx}
\usepackage{braket}
\usepackage{epstopdf}
\usepackage{hyperref}
\usepackage{float}
\usepackage{xcolor}
\usepackage{tikz}
\usetikzlibrary{positioning}
\usepackage{graphicx}
\usepackage{dcolumn}
\usepackage{bm}
\usepackage[normalem]{ulem}
\usepackage{amsmath,amssymb}
\usepackage{subfig}
\usepackage{graphicx}
\usepackage{braket}
\usepackage{epstopdf}
\usepackage{hyperref}
\usepackage{float}
\usepackage{xcolor}
\usepackage{tikz}
\usetikzlibrary{positioning}

\begin{document}
\newcommand{\R}[1]{\textcolor{red}{#1}}
\newcommand{\B}[1]{\textcolor{blue}{#1}}. 

\title{Imbalance for a family of one-dimensional incommensurate models with mobility edges}


\author{Sayantan Roy}
\email{roy.369@buckeyemail.osu.edu }
\affiliation{
Department of Physics, The Ohio State University, Columbus, USA}
\affiliation{Department of Physics, Indian Institute of Science, Bangalore, India}
\affiliation{International Centre for Theoretical Sciences, Tata Institute of Fundamental Research, Bangalore, India}

\author{Subroto Mukerjee}
\email{smukerjee@iisc.ac.in}
\affiliation{Department of Physics, Indian Institute of Science, Bangalore, India}

\author{Manas Kulkarni}\email{manas.kulkarni@icts.res.in}
\affiliation{International Centre for Theoretical Sciences, Tata Institute of Fundamental Research, Bangalore, India}
\begin{abstract}
 In this paper, we look at four  generalizations of the one dimensional Aubry-Andre-Harper (AAH) model which possess mobility edges. We map out a phase diagram in terms of population imbalance, and look at the system size dependence of the steady state imbalance. We find non-monotonic behaviour of imbalance with system parameters, which contradicts the idea that the relaxation of an initial imbalance is fixed only by the ratio of number of extended states to number of localized states. We propose that there exists dimensionless parameters, which depend on the fraction of single particle localized states, single particle extended states and the mean participation ratio of these states. These ingredients fully control the imbalance in the long time limit and we present numerical evidence of this claim. Among the four models considered, three of them have interesting duality relations and their location of mobility edges are known. One of the models (next nearest neighbour coupling) has no known duality but mobility edge exists and the model has been experimentally realized. Our findings are an important step forward to understanding non-equilibrium phenomena in a family of interesting models with incommensurate potentials. 
\end{abstract}

\maketitle

\tableofcontents
\section{Introduction}

 It is a well known fact that for arbitrarily small random disorder, the single particle eigenstates of a non interacting Hamiltonian are all localized in $d < 3$. \cite{PhysRev.109.1492}. This phenomenon, termed Anderson Localization, has been observed in a wide variety of systems\cite{418ca562a1bb4876824c06059f5b12c7,aspect2009anderson}. For $d = 3$, there is a critical value of disorder, above which a localization transition occurs, with a sharp energy dependent mobility edge.\cite{semeghini2015measurement} 

However, after the seminal work by Anderson, it was shown by Andre and Aubrey that localization can occur even in the presence of a quasi-periodic potential \cite{aubry1980analyticity}. Their model is described by the Hamiltonian
\begin{equation}
 H = \sum_i \left[ t( c^\dagger_{i+1} c_i + \rm{h.c.}) + \lambda \cos( 2\pi \beta i+\phi) \right]
\end{equation}
where particles are annihilated (created) by the operators $c_i$ ($c^\dagger_{i}$) and they hop with amplitude $t$ on a one dimensional lattice whose sites are labeled by the index $i$. $\lambda$ is the strength of the on-site potential  and $\beta$ is an irrational number, which ensures the incommensurability of the potential with the lattice. This Hamiltonian also arises in the context of the quantum Hall effect on a lattice with an irrational flux per plaquette \cite{harper1955general,hofstadter1976energy}.  For the AAH Model, there exist a critical disorder strength, below which {\em all} single particle eigenstates are delocalized, and above which, {\em all} single particle eigenstates are localized, which is in contrast to the phenomenon of Anderson localization in one dimension where all single particle states are localized.

The AAH model possesses a duality which maps its localized and delocalized phases onto each other with the transition between them appearing at a self-dual point. The self dual point occurs at $\lambda = 2|t|$, where the eigenstates are neither localized nor delocalized, but  are critical\cite{ostlund1983one}. Transport is sub-diffusive at the critical point \cite{purkayastha2018anomalous}.  The above type of quasi-periodic potential has recently been realized in optical lattices in the context of the experimental observation of Many Body Localization\cite{schreiber2015observation,luschen2017observation,luschen2017signatures} and topologically protected edge states\cite{kraus2012topological,kraus2012topological2}. Variants of this model which contain a single particle mobility edge has also been realized experimentally\cite{luschen2018single,kohlert2019observation,an2020observation}.
\\

 The AAH model can be modified to produce single particle mobility edges, and this has been studied recently in a wide context theoretically\cite{wang2019localization,wang2020one,li2015many,modak2015many,ganeshan2015nearest,sun2015localization,johansson2015comment,deng2017many,nag2017many,biddle2009localization,biddle2010predicted,li2017mobility,li2016quantum,gopalakrishnan2017self}.  In fact, it is believed that a generic one dimensional model with a quasi-periodic potential possesses single particle mobility edges and their lack in the AAH model is due to its specific (fine-tuned) form\cite{luschen2018single,wang2019localization,szabo2020mixed}. 
 
 It is paramount to mention that one of the major advantages of studying such models  is that they mimic certain aspects of the behavior of generic disordered systems in three dimensions, such as the presence of a mobility edge while affording the computational advantage of being in one dimension.

It is well established that the properties of delocalization and localization are deeply connected to those of a measure of ergodicity and non ergodicity respectively in interacting systems\cite{de2014anderson,baum2019avoiding,basko2006problem}. Ergodicity ensures that at long times the system with any memory of its initial conditions settles into an equilibrium state. Generically, this equilibrium  state does not possess broken translational order (Note that the equilibrium  state under discussion is a generic one at any energy rather than say,  the ground state, which is special and can have specific types of order). Therefore, one promising method to determine whether a given system is ergodic is to start it  out in a state with broken translation order and observe whether that order disappears at long times. This  is the method employed in several experimental and theoretical studies of Many-Body Localization via introducing the concept of imbalance as a diagnostic. For  the lattice systems we consider here,  the imbalance $I(t)$ at anytime $t$ as a measure of this order and is defined as $I(t) = \frac{n_{e}(t)-n_{o}(t)}{L}$, where $n_{e}(t)$ and $n_{o}(t)$ are the total  number of particles at even and odd sites respectively and $L$ is the size of the system. 

For an ergodic system, an  initially non-zero value of $I(t)$  is expected to go to zero  at long times, while for a non-ergodic  one, it remains non-zero. Since, non-ergodicity  is  an essential feature of any localized  system, this imbalance of a localized system, even one without interactions, is expected to remain non-zero. However,  the situation for non-interacting systems with delocalized states (regardless of whether they occupy the entire energy spectrum or only a part of it) is  less  obvious since these systems are  not  strictly ergodic (not possessing  interactions). We investigate the behavior of the imbalance for different types of localized, delocalized and ``mixed'' systems in this work.

 In this paper, we calculate the steady state imbalance for various non-interacting models with quasi periodic potentials, obtained by modifying the AAH Model. The calculations are performed by numerically solving the Schr\"{o}dinger equation on finite-sized systems and the imbalance is obtained as a function of the various microscopic parameters and the length $L$ of the system. We perform finite-size scaling to determine the value of the imbalance in the thermodynamic limit and its dependence on system size. Further, we examine its variation with  the microscopic parameters of the models under consideration  and  we find  a non-monotonic dependence on them. We show that the value of the imbalance is not simply determined by the relative fraction of localized to delocalized states but rather also depends on the strength of localization and delocalization of these states. This is quantified by the measures $\epsilon$ and $\epsilon'$ (to be defined later) whose values we show are correlated with that of the imbalance. In other words, in addition to how many localized or de-localized states exist we demonstrate that it is also of paramount importance to quantify how ``localized" a localized state is and how ``de-localized" a delocalized state is.

The rest of the paper is organized as follows.  In section \ref{Models_theory} we introduce the models (see also Table.~\ref{Models}), and point out their main features. We present a brief derivation of the exact mobility edge for the first two generalized AAH models, following Ref~ \onlinecite{ganeshan2015nearest}, and for the slowly varying deterministic potential model from Ref.~\onlinecite{sarma1990localization}.  Section \ref{Calculations} is dedicated to describe the methods employed and calculations.  

 In section \ref{Data}, we calculate the fraction of localized states, whenever the exact expression for the mobility edge is known, and use it to construct the phase diagram for the model. In case the location of the mobility edge is not known, we use the mean participation ratio (which is introduced in section \ref{Models_theory}, Model 3) to map out an approximate phase diagram. We also look at the evolution of a broken translation order in terms of  a charge density wave (CDW), and calculate the imbalance in the long time limit. The long time value of imbalance serves as an order parameter, and can be used to construct an approximate phase diagram, akin to the fraction of localized states. For the main part of this paper, we focus on calculations of Model 1 only, and show that all the justifications for the behaviour of Model 1 in different phases with respect to this order parameter may be extended to the other three models in the Appendix. We look at finite size effects, and propose a leading order scaling of the imbalance with system size. This is used to extract the thermodynamic limit of the imbalance, $I_{0}$ which unveils a non trivial behaviour across the parameter space, noted first in Model 1 in Ref.~\onlinecite{purkayastha2017nonequilibrium}. We extend this analysis for the other three models as well, and show that the non monotonicity is apparent in them as well.

In this same section (\ref{Data}), we provide an explanation to this counter intuitive behaviour by using two dimensionless parameters, $\epsilon$ and $\epsilon'$, following a recent application of them to distinguish between ergodic and MBL phase \cite{modak2018criterion}. These parameters capture the effective strength of the single particle localized states versus extended states in confining the particles' motion, and hence, are directly linked to the behaviour of imbalance as a function of microscopic parameters of the model. In Section \ref{conclusion}, we summarize our results along with a brief outlook. 
A discussion of the other three models are presented in the appendix, with appendix Section \ref{FSS_appendix} discussing the system length dependence of the steady state imbalance and it's behaviour in different phases. 
 Appendix section \ref{Mobilityedges_M3_appendix} shows how the appearance
of multiple mobility edges show a unique transition between phases in Model 3.

\section{Models}\label{Models_theory}

\begin{table*}[t]
\centering
\begin{tabular}{|c|c|c|c|}
\hline
     Model & Hopping term & Onsite Potential $V_i$ & Location of Mobility Edge\\
     \hline
     1 & $T_{i,i+1}$ = $t (a^{\dagger}_{i}a_{i+1}+hc)$& $ 2\lambda \frac{ \cos (2\pi \beta i +\phi)}{1-\alpha \cos(2\pi \beta i + \phi)}a^{\dagger}_{i}a_{i}$ & $\alpha E_{c} = 2 sgn(\lambda)(|t| - |\lambda|)$ \\
     \hline
     2 & $T_{i,i+1}$ = $t(a^{\dagger}_{i}a_{i+1}+hc)$& $  2\lambda\frac{(1- \cos(2\pi \beta i +\phi)}{1+\alpha \cos(2\pi \beta i + \phi)}a^{\dagger}_{i}a_{i}$ & $ \alpha E_{c} = 2 sgn(\lambda)(|t| - |\lambda|)$ \\
     \hline
     3 & $T_{i,i+1}$ = $t(a^{\dagger}_{i}a_{i+1}+hc)$& $  2\lambda \cos(2\pi \beta i +\phi)a^{\dagger}_{i} a_{i}$ & No analytical expression,\\ 
     & $T_{i,i+2}$= $\alpha t(a^{\dagger}_{i}a_{i+2}+hc)$& &phases calculated through PR \\
     \hline
     4& $T_{i,i+1}$ = $t(a^{\dagger}_{i}a_{i+1}+hc)$& $ 2\lambda \cos(2\pi \beta i^{\alpha} + \phi)a^{\dagger}_{i}a_{i}$ & $E_{c} = \pm 2(|t| - |\lambda|)$ \\
     \hline
\end{tabular}
\caption{Models studied and their properties. All the models feature delocalized, intermediate phase with mobility edge and localized phases, determined by the parameters $\lambda, \alpha$. [Figs.~\ref{Phase_diagram_M1}(a) - \ref{Phase_diagram_M4}(a)]}
\label{Models}
\end{table*}

All the  models we  study contain an on-site potential $V_i$ of strength $\lambda$. The form of the potential also contains an irrational number $\beta$  (which we set equal to the golden mean $\frac{\sqrt{5}+1}{2}$) that ensures its incommensurability with the underlying lattice and a phase $\phi$ which can  be used to move the potential relative to the lattice. $\phi$ is averaged over in  our calculation analogous to disorder averaging. Finally, the potential also depends on an  auxiliary parameter $\alpha$, which can be tuned to move the  position  of the mobility  edge in the spectrum. The models studied are listed in Table \ref{Models} along  with some of their specific properties. We restrict to $\lambda \in [0,2]$ and $\alpha \in [0,1]$ unless otherwise specified. 

The first two of the four models we study were explored by Ref \onlinecite{ganeshan2015nearest}. The models were shown to be self dual by a generalized transformation akin to a Fourier transformation, and unlike the AAH model, have mobility edges due to energy dependent self dual conditions.  The first of this family of self dual models is the Generalized AAH Model (GAAH), described by:

\begin{equation}
H^{(1)} = t\sum_i (a^{\dagger}_{i}a_{i+1}+hc) + \sum_{i} \frac{2\lambda \cos(2\pi \beta i +\phi)}{1-\alpha \cos(2\pi \beta i + \phi)}a^{\dagger}_{i}a_{i} 
\label{GAAH 1}
\end{equation}

Another modification to this model, which preserves the self duality condition, and transforms under the same transformation \cite{ganeshan2015nearest} is

\begin{equation}
H^{(2)} = t\sum_i (a^{\dagger}_{i}a_{i+1}+hc) + 2\lambda \sum_{i} \frac{1- \cos(2\pi \beta i +\phi)}{1+\alpha \cos(2\pi \beta i + \phi)}a^{\dagger}_{i}a_{i} 
\label{GAAH 2}
\end{equation}

The Schor\"{o}dinger equation for both the models are written out, following Ref.~\onlinecite{ganeshan2015nearest} ($g$ below contains the details of the specific model)

\begin{equation}
    t(\psi_{i-1}+\psi_{i+1})+ g\chi_{i}(\delta)\psi_{i} = (E+2\lambda \cosh \delta)\psi_{i}
    \label{SE}
\end{equation}

with the definition of the onsite potential $\chi_{i}$,\cite{ganeshan2015nearest}
\begin{equation}
    \chi_{i}(\delta) = \frac{\sinh(\delta)}{\cosh(\delta) - \cos(2\pi\beta i + \phi)}. 
\end{equation}
where $\delta$ is defined in Eq.~\ref{delta}. The location of the mobility edge can be found using the self-dual transformation of the amplitude of the wavefunction at site $i$, $\psi_{i}$  to amplitude at point $k$ in the dual space \cite{ganeshan2015nearest} 
\begin{equation}
f_{k} = \sum_{mni} e^{i2\pi b(km + mn + ni)} \chi_{n}^{-1}(\delta_{0})\psi_{i}
\label{transform}
\end{equation}
which transforms the Schr\"{o}edinger equation (Eq.~\ref{SE}) to 

\begin{equation}
    t(f_{k-1}+f_{k+1})+g\frac{\sinh \delta}{\sinh \delta_{0}} \chi_{k}(\delta_{0}f_{k}) = 2t\cosh\delta f_{k},
    \label{dual}
\end{equation}

with the parameter $\delta_{0}$ defined as

\begin{eqnarray}
    \delta_{0} &=& \cosh^{-1}\bigg( \frac{E+2\lambda \cosh \delta}{2t}\bigg)\\
    \delta &=& \cosh^{-1}(1/\alpha)
    \label{delta}
\end{eqnarray}

The parameter $g$ is model dependent, and for the model in Eq.~\ref{GAAH 1} is defined as 
$$g= 2\lambda \cosh(\delta)/\tanh(\delta),\quad \text{(Model 1)
}$$  and as 
$$g =2\lambda (1+\cosh \delta)/ \sinh (\delta), \quad\text{(Model 2)}$$ 
for the model in  Eq.~\ref{GAAH 2}.
The self dual condition of Eq.~\ref{SE} to Eq.~\ref{dual} is met by $\delta = \delta_{0}$ which gives the equation for the position of the mobility edge:

\begin{equation}
\alpha E = 2 \text{sgn}(\lambda)(|t| - |\lambda|),\quad\text{(Models 1 and 2})
\label{Dualtiy GAAH}
\end{equation}

For both models, states with energy $E > E_{c}$ are extended, and those with $E <E_{c}$ are localized. Both of these models are a subset of class of models with onsite term as in Eq.~\ref{SE} which may be obtained with an arbitrary choice of the parameters $g(\alpha,t)$ and $E(\lambda, \alpha,t)$ but still have an exact mobility edge. The parameter $\alpha$ is a generalization of various special cases: For Model 1,(Eq.~\ref{GAAH 1}), $\alpha = 0$ produces the AAH limit. For Model 2(Eq.~\ref{GAAH 2}), $\alpha = 0$ corresponds to the AAH Model, $\alpha = -1$ produces the constant onsite potential model with $V_{i} = -2\lambda$ and $\alpha = 1$ produces the closed form singular potential\cite{ganeshan2015nearest} $V_{i} = \tan^{2}\big (\frac{2\pi\beta i + \phi}{2} \big )$.

Another perturbation to the self duality of the AAH model may be introduced by considering long range hopping, as studied in Ref. \onlinecite{biddle2011localization}. The Schr\"odinger equation, Eq.~\ref{SE} becomes

\begin{equation}
    \sum_{i \neq j}t e^{-p|i-j|}\psi_{j} + 2 \lambda \cos(2 \pi \beta i +\phi)\psi_{i} = E \psi_{i} 
    \label{SE Nonnearest}
\end{equation}

where $0<p<1$. The inclusion of an exponentially decaying hopping amplitude  perturbs the AAH duality to the following linear self dual relation 

\begin{equation}
    \cosh(p) = \frac{E+t}{2\lambda}
    \label{Duality NNN}
\end{equation}

In our work we consider only a next nearest neighbour hopping term as perturbation to the AAH model, also called as the $t_{1}- t_{2}$ model. Without loss of generality, we parameterize the hopping amplitudes as $t_{1} = t, t_{2} = \alpha t$.  In this parameterization, consider

 \begin{eqnarray}
H^{(3)} &=& t\sum_i (a^{\dagger}_{i}a_{i+1}+ \alpha a^{\dagger}_{i}a_{i+2}+hc) \nonumber \\ &+& \sum_{i} 2\lambda \cos(2\pi \beta i +\phi)a^{\dagger}_{i}a_{i} 
\label{nnn Model}
\end{eqnarray} 

This model produces the nearest neighbour hopping AAH Model with $\alpha = 0$, and two superimposed AAH lattice with nearest and next nearest hopping at $\alpha = 1$. This model does not have an exact expression for mobility edge, although studies have shown presence of both localized and delocalized states coexisting at the same value of the microscopic parameters \cite{biddle2011localization,biddle2009localization}.

A first order approximation to the hopping amplitude in Eq.~\ref{SE Nonnearest} may be made by considering all hoppings longer than the next nearest neighbour interaction to be suppressed, by a small value of p. An approximate relation for the location of mobility edge is constructed by substituting $ p = \ln (t_{1}/t_{2})$ into Eq.~\ref{Duality NNN} which yields

\begin{equation}
    \alpha + \frac{1}{\alpha} \sim \frac{E+t}{\lambda}
    \label{nnn root}
\end{equation}

and this gives 

\begin{equation}
E_{c} \sim \lambda(\alpha^{2} + 1)/\alpha - t,\quad \text{(approximate for Model 3)}
\end{equation}

As found in Ref.~\onlinecite{biddle2011localization}, this works only for small values of $p$, or large values of $\alpha$, as shown by Inverse Participation ratio calculations.  To explore the phases in this model, an exact diagonalization of the Hamiltonian is employed to get the eigenstates and calculate the Participation ratio (PR) of the $n^{\text{th}}$ eigenstate, defined as

\begin{equation}
    PR(\psi_{n}) = \frac{1}{\sum_{j}|\psi_{n}(j)|^{4}}
    \label{Part}
\end{equation}

where $\psi_{n}(j)$ is the amplitude of the $n^{\text{th}}$ eigenstate at site j. This is of $O(1)$ for the localized states, and $O(L)$ for delocalized states, $L$ being the system size. Owing to lack of expression for the mobility edge for Model 3, the mean participation ratio (MPR),  which is the PR averaged over all eigenstates $n$ , $MPR = \langle PR \rangle_{ \{n \}}$ will be used to map out a phase diagram for this model, instead of using the fraction of localized states, as done for the other models.

The next model we study contains a slowly varying deterministic potential in the onsite term. These class of models were studied in Ref.~\onlinecite{sarma1990localization} and its geometrical properties without referring to tight binding hamiltonians were explained in Ref \onlinecite{berry1988renormalisation}. The Hamiltonian is

\begin{equation}
H^{(4)} = t\sum_i (a^{\dagger}_{i}a_{i+1}+hc)\\ + \sum_{i} 2\lambda \cos(2\pi \beta i^{\alpha} +\phi)a^{\dagger}_{i}a_{i} 
\label{varying Model}
\end{equation}

The model transitions smoothly from the constant uniform potential at $\alpha = 0$ to the AAH limit at $\alpha = 1$ . For $\alpha > 2$, the potential is ``pseudo random" in the sense that the localization length for the $\alpha > 2$ case was shown to be the same as that in the corresponding random case \cite{sarma1990localization},  and this model can be identified with the 1D Anderson model\cite{sarma1990localization}. For $1<\alpha<2$, it was proved that the eigenstates at the centre of spectra are all localized, but with large localization length, or vanishing Lyapunov exponents \cite{thouless1988localization}. The localization mechanism is different from the Anderson mechanism, as the potential is neither aperiodic, nor random, but is deterministic. Indeed, it was shown that the density of states have discontinuity at the mobility edges, unlike the 3D Anderson model \cite{sarma1990localization}.
The slowly varying nature of the potential is written as

\begin{equation}
    \frac{dV_{i}}{di} = \frac{-4\lambda \pi \beta \cos(2\pi \beta i^{\alpha} +\phi)}{i^{1-\alpha}}
\end{equation}
 which implies that $(V_{i+1}-V_{i}) \propto i^{\alpha-1}$, and this in the limit of large system sizes, goes to zero, as $ 0 < \alpha <1 $.  Note that this almost constant feature of potential is crucial for the localization transition in this model.  The Schr\"odinger equation is solved with the ansatz, $\psi_{n} = z^{n}$, and it may be proved that it is possible to have localized states (extended states) for real $z$ (imaginary $z$) only at

\begin{eqnarray}
\Re(z) = 0 \implies |E| > 2(|t|-|\lambda|)\\
    \Im(z) = 0 \implies |E| < 2(|t|-|\lambda|)
    \label{Amplitudes_Svm}
\end{eqnarray}

 which gives us the location of the mobility edge\cite{sarma1990localization}  (independent of $\alpha$) at 
\begin{equation}
E_{c} = \pm 2(|t|-|\lambda|),\quad \text{for Model 4}
\label{Duality_SV}
\end{equation}  

 All states at the centre of the spectra, with energy $E$, $ |E| < |E_{c}| $ are extended, and those at the tails with $|E| > |E_{c}|$ are localized (for the positive $(\lambda,\alpha) $ quadrant).  


\section{Methods and Calculations}\label{Calculations}

 We employ an exact diagonalization of the Hamiltonians in Eq.~\ref{GAAH 1} (Model 1), Eq.~\ref{GAAH 2} (Model 2), Eq.~\ref{nnn Model} (Model 3), Eq.~\ref{varying Model} (Model 4), with the parameters limited to $\lambda \in [0,2], \alpha \in (0,1)$.  We perform calculations with lattice size up to $L=900$, and use a disorder averaging over $\phi$ to restore translation in variance. All the results shown below are for $\beta = \frac{\sqrt{5}+1}{2}$ (golden mean) unless mentioned explicitly to be a different number. The imbalance is calculated by looking at the number density at each site, from the single particle wave function amplitude, $|\psi_{n}(i)|^{2}$. The imbalance is obtained by summing over the amplitudes, over entire single particle spectrum. The steady state value of the imbalance is calculated from the average in the time window 200-400 $\tau$ where $\tau = \hbar/|t|$ is the unit of time. The imbalance is constant in this interval , with only small fluctuations. 

The single particle wave functions can also be grouped into extended/localized category by looking at their energy and the respective mobility edge expressions for the models,  where it is known analytically. We also verify the distribution by looking at the PR calculation 
(not presented here), defined in Eq.~\ref{Part}. For Model 3 , where the exact mobility edge is not known,  the PR is calculated to differentiate between the single particle states, where it scales as
 
\begin{eqnarray} \label{PRcases}
PR \sim
\begin{cases}
& O(L),\quad \text{de-localized} \\
& O(1),\quad \text{localized} \\
& O(L^{D(q)}),\quad \text{critical} 
\end{cases}
\end{eqnarray}
 
where $D(q)$ is the fractal exponent. The phase diagram is then constructed by looking at the fraction of  localized states, $\eta$ (Model 1,2,4) and MPR (Model 3) for all possible ($\lambda,\alpha$) . We also plot the steady state imbalance over  the parameter space for all models and see that it maps out an approximate phase diagram similar to the one obtained by using the diagnostic $\eta$. 

After a finite size scaling for the steady state imbalance, the imbalance follows a linear scaling in $1/L$ at leading order, whenever the steady state value is high enough to differentiate it from statistical noise. We propose that this is the behaviour at leading order at least in the mobility edge and localized edge phase.  At the critical point, we numerically find that it shows $\frac{1}{L}$ even for relatively small system sizes. This allows us to write the steady state imbalance

\begin{equation}
    I(L) = \frac{a}{L} + I_{0}+ \text{higher order terms}
    \label{L scale}
\end{equation}

The behaviour of $I_{0}$, the thermodynamic limit of the imbalance, immediately captures the trend seen in the imbalance phase diagram, for finite sizes. The behaviour of $I_{0}$ is explained by calculating the two dimensionless parameters, $\epsilon$ and $\epsilon'$ as
 
\begin{eqnarray}
\label{epsilon}
\epsilon &=& \Tilde{\eta} \frac{1 - MPR_{D}/L}{MPR_{L}-1},\,{\text{(degree of delocalization)}}\\
\epsilon{'} &=& \frac{1}{MPR_{L}-1},\,{\text{(degree of localization)}}
\label{epsilonp}
\end{eqnarray}

The parameter $\epsilon$ was used by Ref \onlinecite{modak2018criterion} to define a criteria for MBL to ergodic phase transition in interacting systems, with a single mobility edge.  The parameter $\epsilon$ satisfies the following
 
\begin{eqnarray} \label{epcases}
\epsilon
& <1,\quad \text{ergodic phase} \\
& >1,\quad \text{MBL phase} 
\end{eqnarray}

Here, $\Tilde{\eta}$ is the ratio of the number of localized states
to delocalized single particle states,
 $MPR_{D}(MPR_{L})$ is the mean participation ratio  averaged over the spectrum and over all $\phi$ disorders following Eq.~\ref{Part} over the delocalized (localized) states only.  Note that $\epsilon$ captures how strongly localized the single particle localized states are versus how strongly delocalized the single particle delocalized states are, only in the mobility edge phase.  \cite{modak2018criterion}.  This is by construction $0$ ($\tilde{\eta} =0$) in the delocalized phase, and $\infty$ (due to Eq.~\ref{PRcases}) in the localized phase. 
 
 To look at the degree of localization of the localized states only, a new parameter $\epsilon'$ is defined in  Eq.~\ref{epsilon}. We show that the parameters $\epsilon$ and $\epsilon'$ are capable of explaining the behaviour of $I_{0}$ for all the models in their various phases. The parameter $\epsilon$ has jump discontinuity unlike $\epsilon'$, whenever the fraction of localized states $\eta$, and hence $\Tilde{\eta}$ change over the phase diagram in the mobility edge phase for all the models.
 
\section{Results}\label{Data}

 In this section, we present the results of our calculations of phase diagram, steady state imbalance, and system size effects for Model 1 (Eq.~\ref{GAAH 1})  in section \ref{Models_theory}, and argue that imbalance provides a richer description of the system and it's phases than a simple fraction of states or PR.

\subsection{Phase Diagram}\label{Phases}

An exact diagonalization is used, as described in the last section, to map out the phase diagrams for all the

 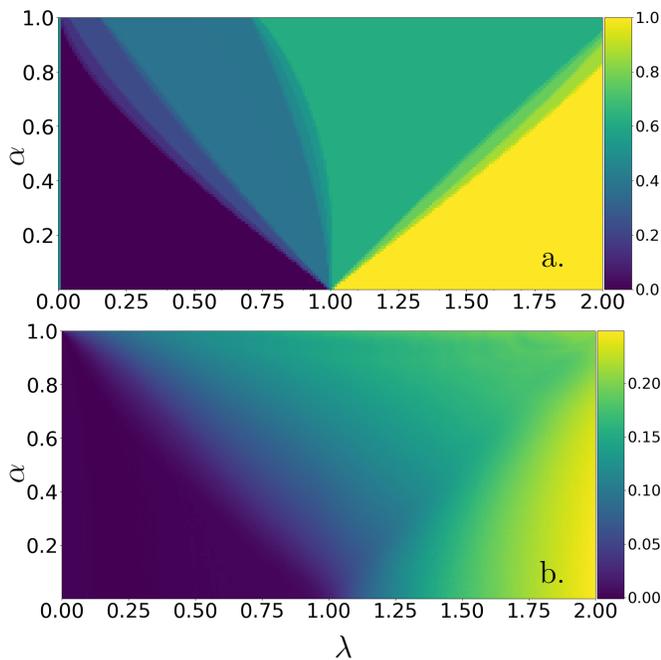
\begin{figure}[H]
\begin{tikzpicture}
\node (img1) {\includegraphics[width=\linewidth]{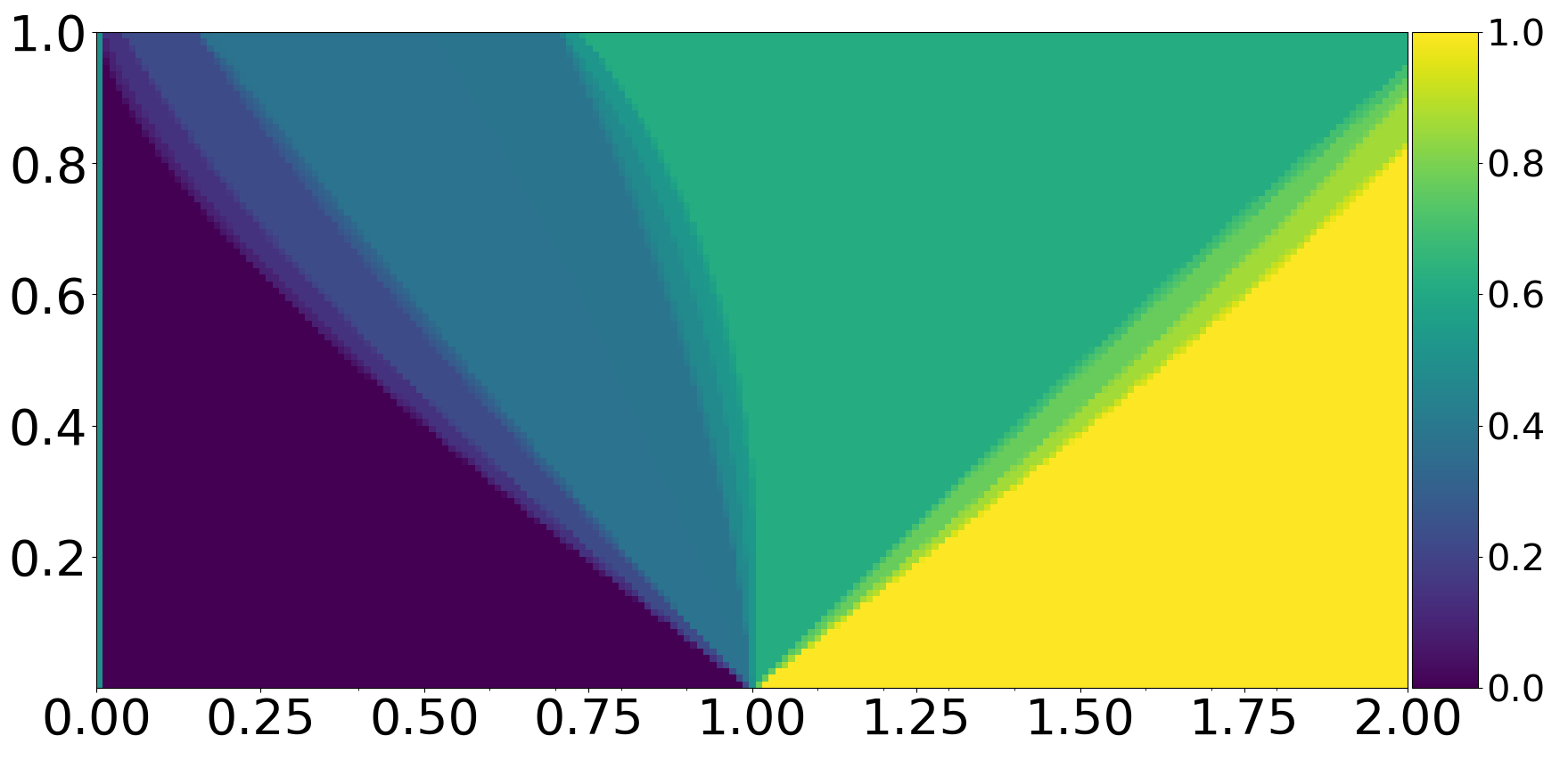}};
\node[left=of img1,node distance=0cm,rotate=90,anchor=center,yshift=-1.1cm]{\large{ $\alpha$}};
\node[left=of img1,node distance=0cm,yshift=-1.3cm,xshift=8.5cm]{\large{a.}};
\node (img2) [below=of img1,yshift=1.35cm]{\includegraphics[width=\linewidth]{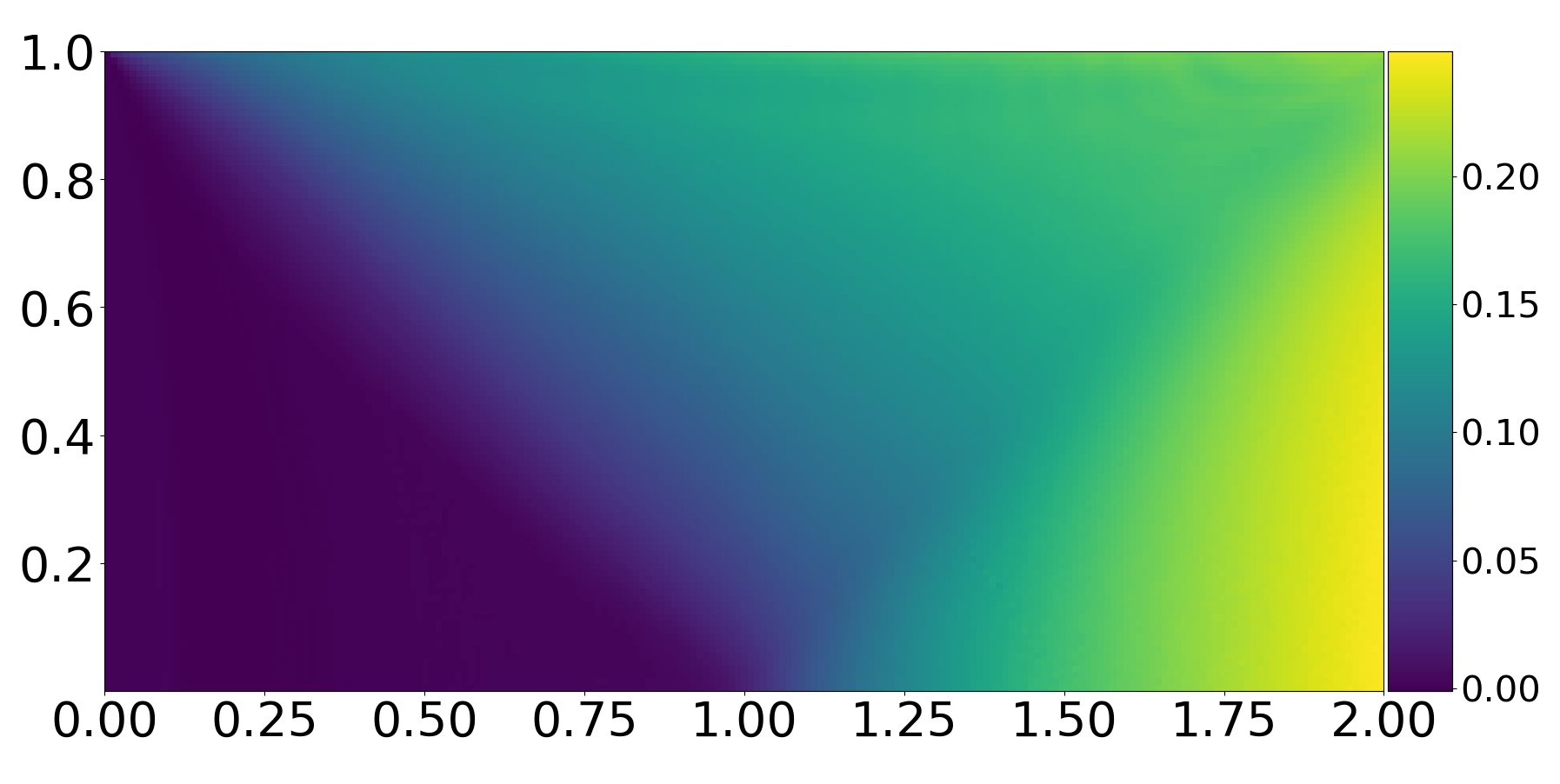}};
\node[left=of img2,node distance=0cm,rotate=90,anchor=center,yshift=-1.1cm]{\large{$\alpha$}};
\node [below= of img2,node distance=0.2cm,yshift=1.25cm]{\large{$\lambda$}};
\node[left=of img2,node distance=0cm,yshift=-1.3cm,xshift=8.5cm]{\large{b.}};
\end{tikzpicture}
\caption{ (a) Phase diagram of Model 1 for $ \beta = \frac{\sqrt{5}+1}{2}$ as a function of $\lambda$ and $\alpha$ in terms of fraction of  localized states $\eta$ (single disorder realization). (b) Steady state imbalance for Model 1 ($L = 128$, averaged over 1000 disorder configurations).} 
\label{Phase_diagram_M1}
\end{figure}
\begin{figure}[H]
\begin{tikzpicture}
\node (img3) {\includegraphics[width=\linewidth]{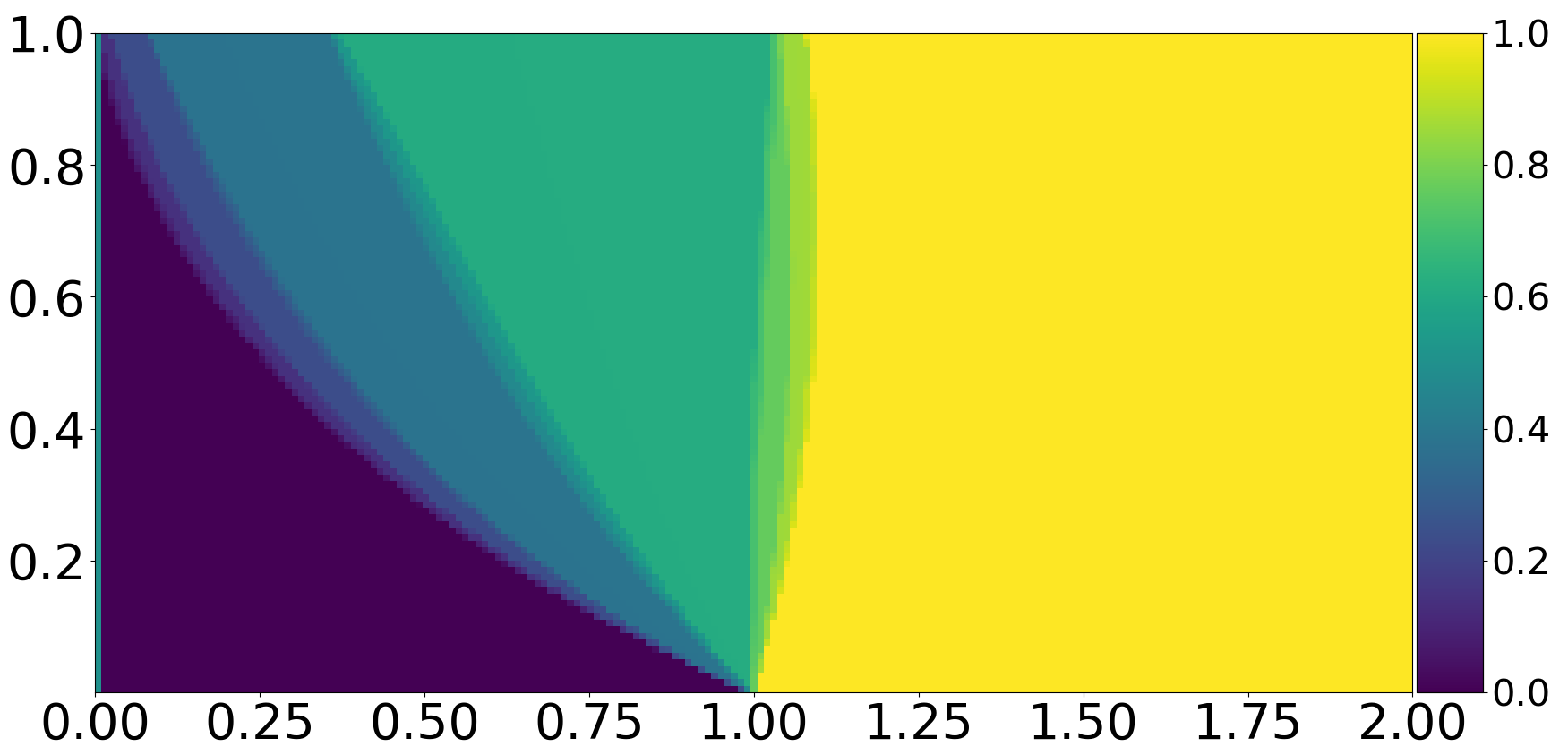}};
\node[left=of img3,node distance=0cm,rotate=90,anchor=center,yshift=-1.1cm]{\large{ $\alpha$}};
\node[left=of img3,node distance=0cm,yshift=-1.3cm,xshift=8.5cm]{\large{a.}};
\node (img4) [below=of img3,yshift=1.25cm]{\includegraphics[width=\linewidth]{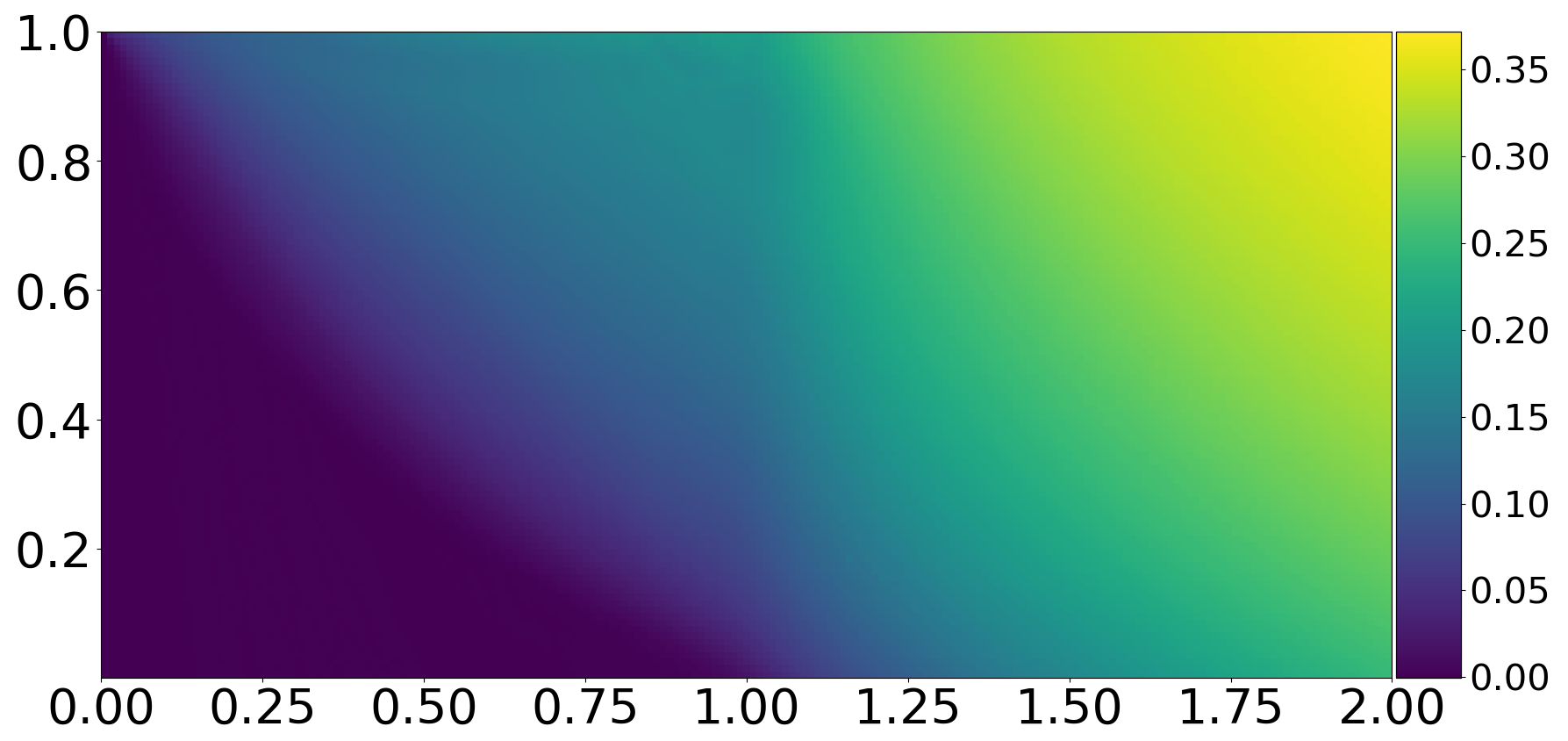}};
\node[left=of img4,node distance=0cm,rotate=90,anchor=center,yshift=-1.1cm]{\large{$\alpha$}};
\node [below= of img4,node distance=0.2cm,yshift=1.25cm]{\large{$\lambda$}};
\node[left=of img4,node distance=0cm,yshift=-1.3cm,xshift=8.5cm]{\large{b.}};
\end{tikzpicture}
\caption{a) Phase diagram of Model 2 for $ \beta = \frac{\sqrt{5}+1}{2}$ as a function of $\lambda$ and $\alpha$ in terms of fraction of localized states $\eta$ (single disorder realization). (b) Steady state imbalance for Model 2 ($L = 256$ sites, averaged over 1000 disorder configurations). }
\label{Phase_diagram_M2}
\end{figure}

\begin{figure}[H]
\begin{tikzpicture}
\node (img5)  {\includegraphics[width=\linewidth]{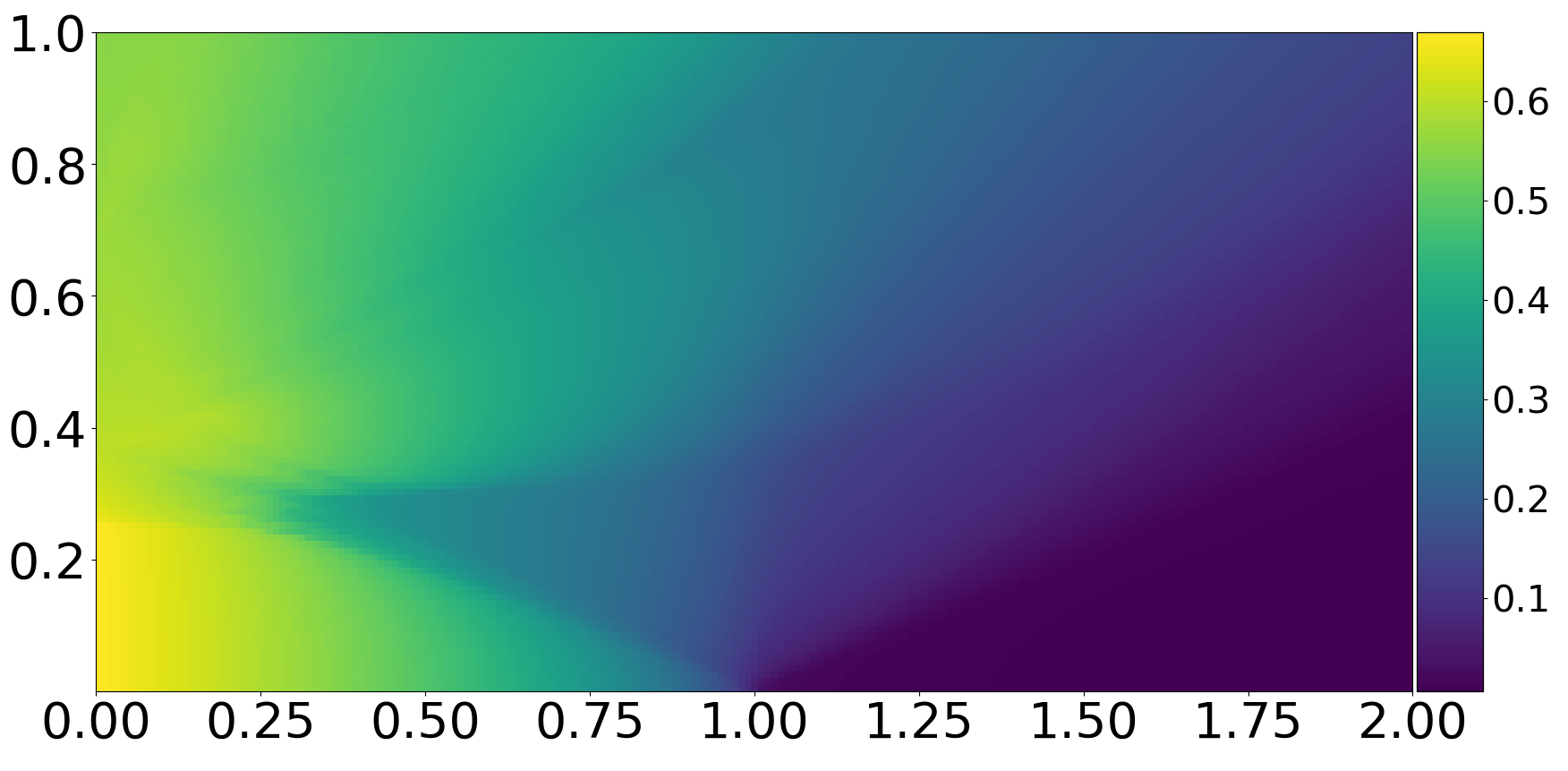}};
\node[left=of img5,node distance=0cm,rotate=90,anchor=center,yshift=-1.1cm]{\large{ $\alpha$}};
\node[left=of img5,node distance=0cm,yshift=-1.3cm,xshift=2.5cm]{\large{a.}};
\node (img6) [below=of img1,yshift=1.25cm]{\includegraphics[width=\linewidth]{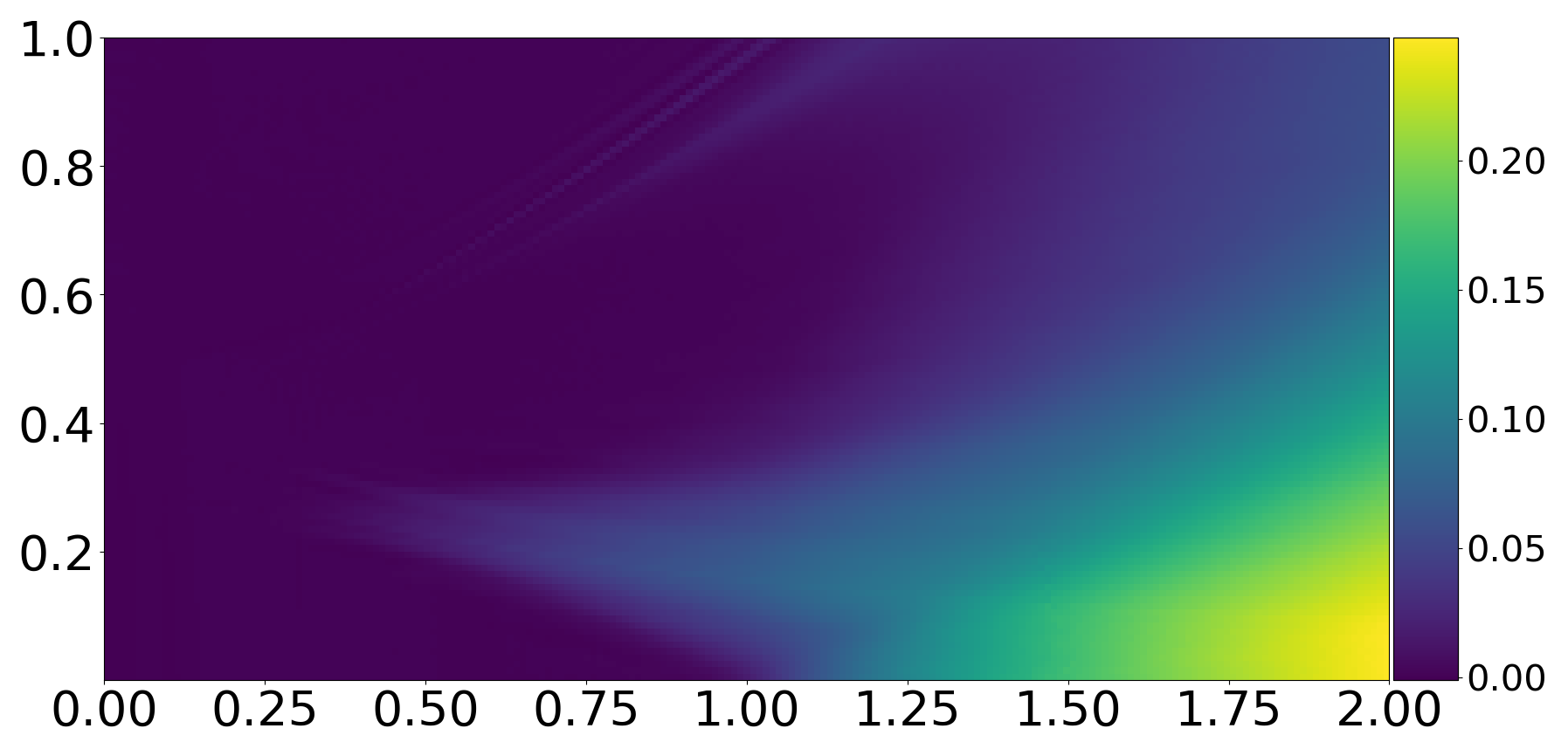}};
\node[left=of img6,node distance=0cm,rotate=90,anchor=center,yshift=-1.1cm]{\large{$\alpha$}};
\node [below= of img6,node distance=0.2cm,yshift=1.25cm]{\large{$\lambda$}};
\node[left=of img6,node distance=0cm,yshift=-1.3cm,xshift=2.5cm,color = white]{\large{b.}};
\end{tikzpicture}
\caption{ a) Phase diagram of Model 3 for $ \beta = \frac{\sqrt{5}+1}{2}$ as a  function of $\lambda$ and $\alpha$ in terms of MPR/$L$. $MPR/L\sim0$ means localized phase , $MPR/L \sim O(1)$ means delocalized phase.  (b) Steady state imbalance for Model 3 ($L$ = 256 sites, averaged over 1000 disorder configurations). }
\label{Phase_diagram_M3}
\end{figure}
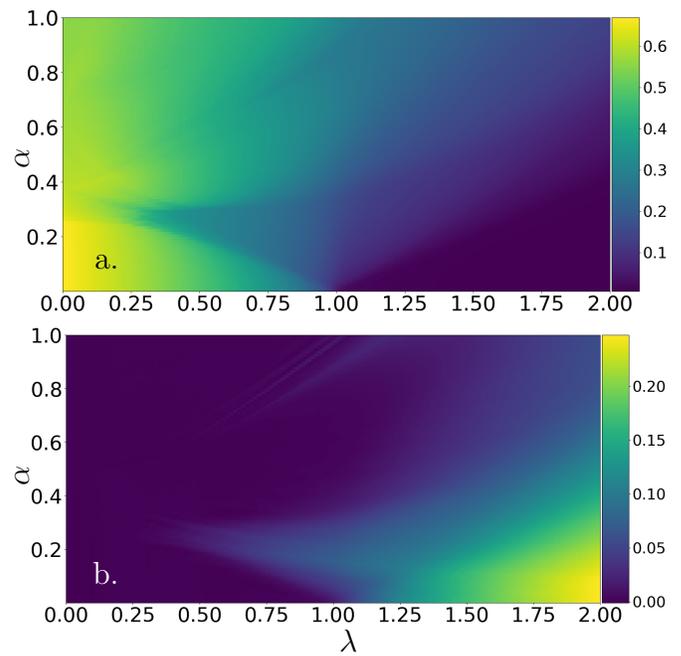
\begin{figure}[H]
\begin{tikzpicture}
\node (img7) {\includegraphics[width=\linewidth]{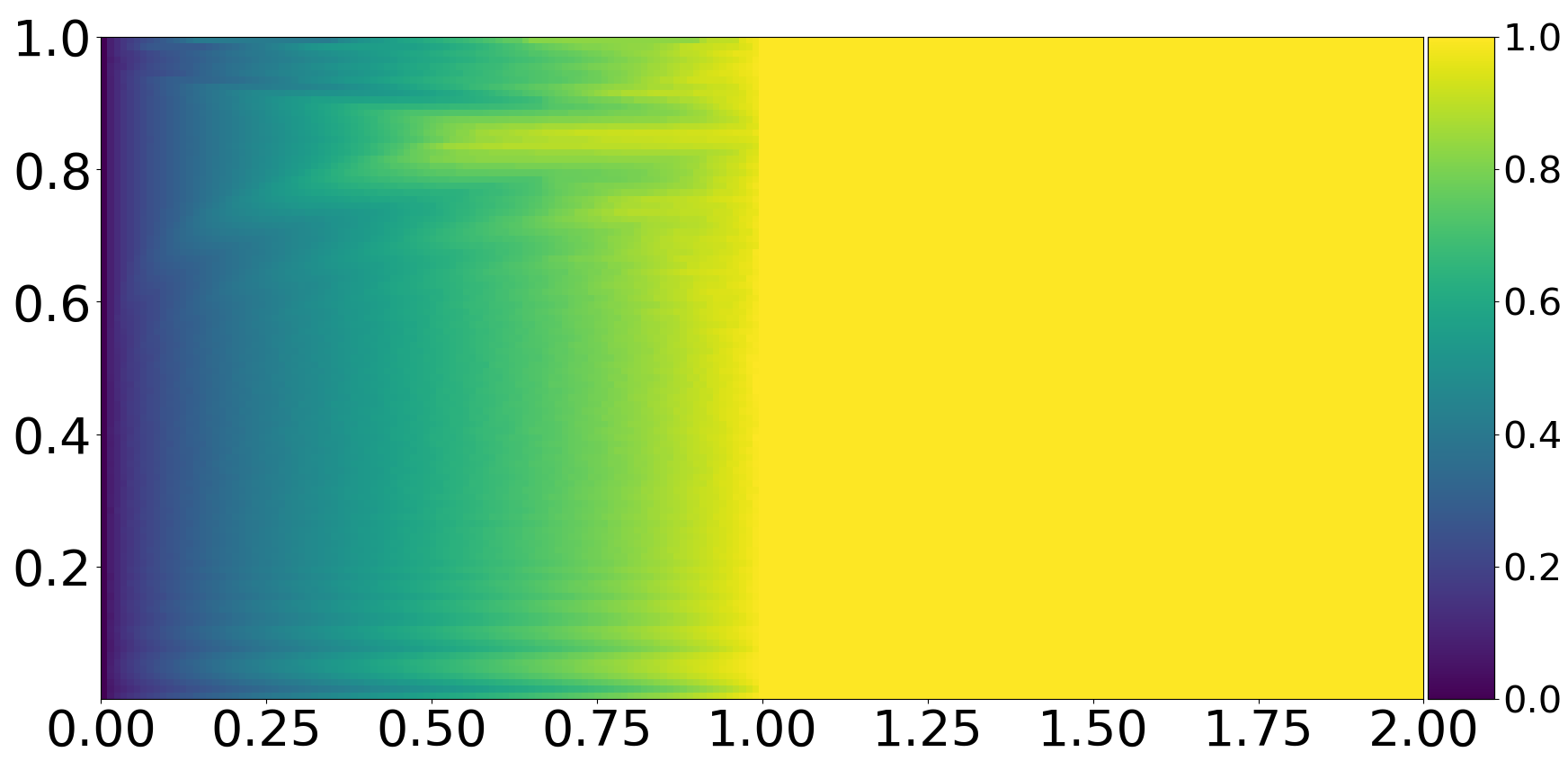}};
\node[left=of img7,node distance=0cm,rotate=90,anchor=center,yshift=-1.1cm]{\large{ $\alpha$}};
\node[left=of img7,node distance=0cm,yshift=-1.3cm,xshift=8.5cm]{\large{a.}};
\node (img8) [below=of img7,yshift=1.25cm]{\includegraphics[width=\linewidth]{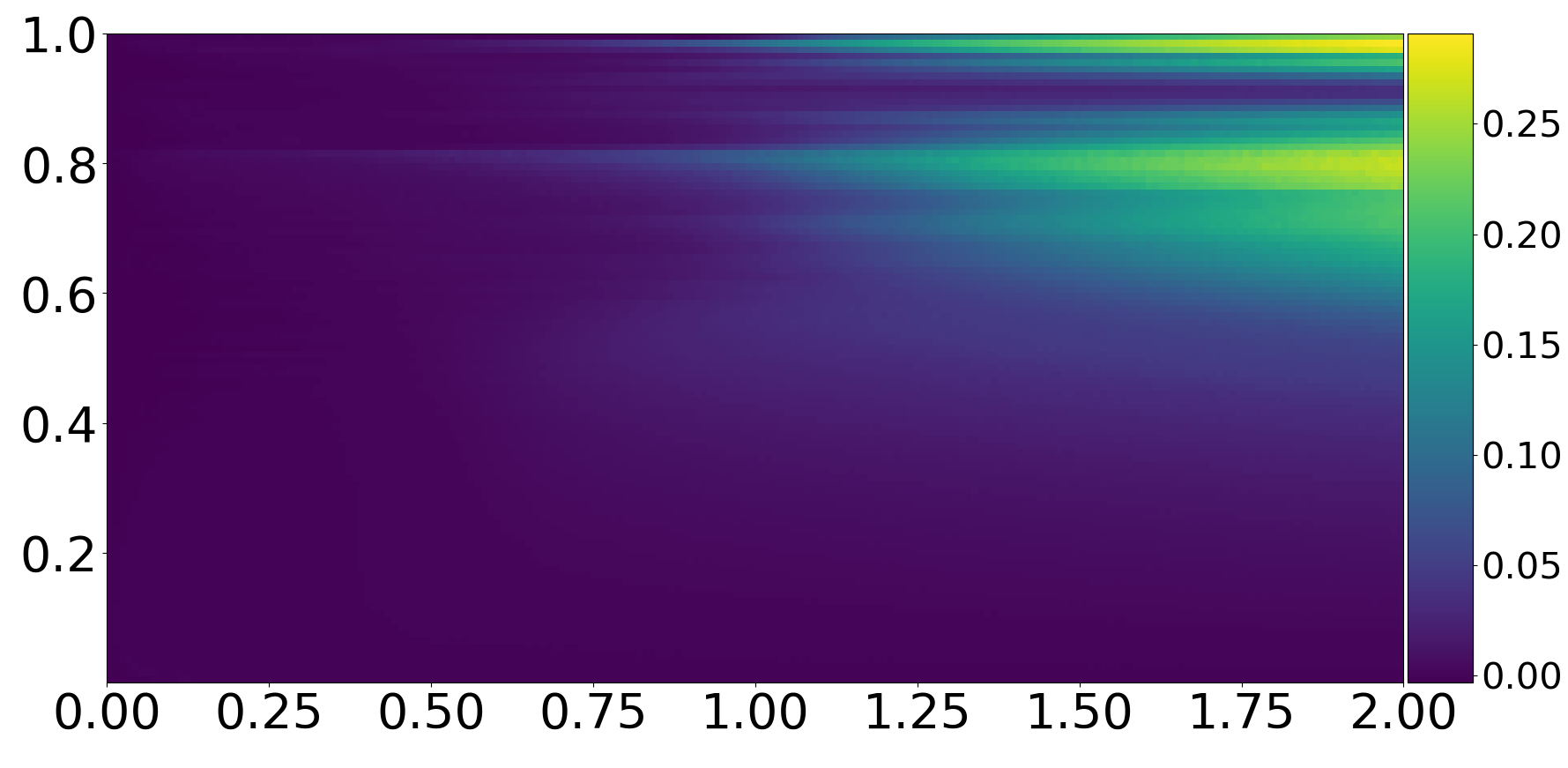}};
\node[left=of img8,node distance=0cm,rotate=90,anchor=center,yshift=-1.1cm]{\large{$\alpha$}};
\node [below= of img8,node distance=0.2cm,yshift=1.25cm]{\large{$\lambda$}};
\node[left=of img8,node distance=0cm,yshift=-1.3cm,xshift=8.5cm,color = white]{\large{b.}};
\end{tikzpicture}
 \caption{a) Phase diagram of Model 4 for $ \beta = \frac{\sqrt{5}+1}{2}$ as a function of $\lambda$ and $\alpha$ in terms of fraction of localized states $\eta$ (single disorder realization). (b) Steady state imbalance for Model 4  ($L = 256$ sites, averaged over 1000 disorder configurations).}
\label{Phase_diagram_M4}
\end{figure}
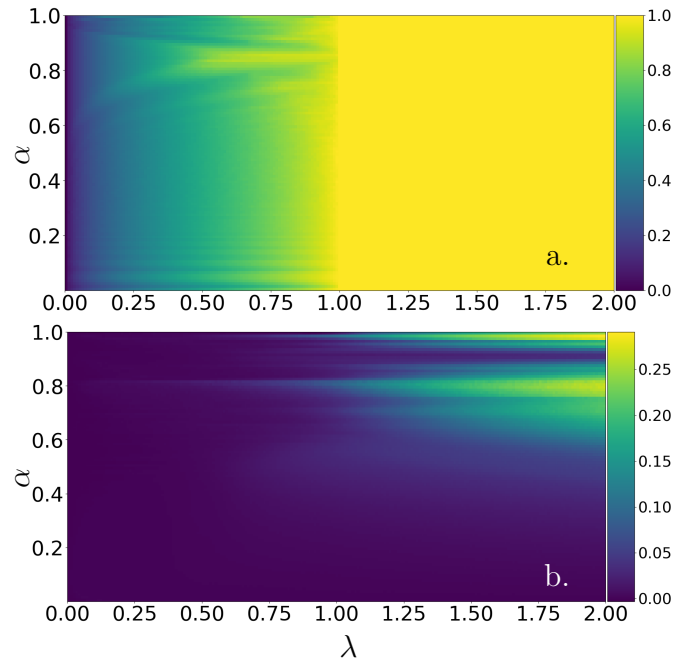

models, listed in section II and Table.~\ref{Models}. These are presented in Fig.~\ref{Phase_diagram_M1}(a) - \ref{Phase_diagram_M4}(a).  For Models 1,2,4, the phase diagram is represented in terms of the fraction of localized states, $\eta$ calculated from Eq.~\ref{Dualtiy GAAH} and Eq.~\ref{Duality_SV} for Model 1,2 and 4, and in terms of the MPR, obtained from Eq.~\ref{Part} for Model 3.  The phase diagram of Model 1 with $\beta = (\sqrt{5}-1)/2$ has been obtained previously \cite{modak2018criterion} and the phase diagram with $\beta = (\sqrt{5}+1)/2$ has been obtained in Ref.~ \onlinecite{purkayastha2017nonequilibrium}. The latter agrees with the one we have obtained and shown in Fig.\ref{Phase_diagram_M1} (a).  

 For Model 2 [Fig.\ref{Phase_diagram_M2} (a) ], the choice of $\beta = (\sqrt{5}+1)/2$ wipes out the mobility edge phase for $\lambda >1$  as seen for Model 1,altogether . However, the localized states for both the models are very weakly localized with large localization length, for small values of $\alpha$ near the AAH critical point.  But with increasing $\alpha$, the localized states in Model 2 are strongly localized (small localization length) and are different from the ones in Model 1 where they are weakly localized (large localization length).   
For Model 3 [Fig.\ref{Phase_diagram_M3} (a) ], the phase diagram reveals a rich transition physics in the region of $\lambda < 1$, where the system transitions from delocalized to mobility edge to back to delocalized phase. This is attributed to appearance and merger of mobility edge. The system departs from the delocalized phase to the mobility edge phase, as $\alpha$ increases from 0, due to appearance of a single mobility edge. At higher values of $\alpha$, multiple mobility edges appear in the system, and they merge together to bring the system back in to the delocalized phase at some higher $\alpha$. This is shown using PR calculation, in Fig.~\ref{IPR_appendix} (see Appendix) for multiple values of $\alpha$ along a fixed $\lambda$ at which this peculiar and remarkable transition happens. 

 Model 4 [Fig.~\ref{Phase_diagram_M4} (a) ]  shows an exclusively localized phase for $\lambda >1$, for both the choices of $\beta$. However, it is also worth noting that the phase diagram is identical for both values of $\beta$, which implies that the spectrum is insensitive to the choice of the irrational number. This is in agreement with the claim that the parameter $\beta$ is an irrelevant parameter for the model, from a scaling theory perspective \cite{sarma1990localization}. 

 Next we discuss the steady state imbalance as a function of parameters of the Hamiltonians of the four models. 
 
\subsection{Imbalance}\label{SSI}
We see that the imbalance can also be used to identify the different phases and thus map out an approximate version of the phase diagram. At long times, it remains close to the initial value of 0.5 deep in the localized phase, asymptotes to 0 deep in the delocalized phase and to intermediate values in the vicinity of the mobility edge. This is shown for all the models in Figs.~ \ref{Phase_diagram_M1}(b)- \ref{Phase_diagram_M4}(b). 
 Results of the previous discussion (which is top panel of Figs.~ \ref{Phase_diagram_M1}-\ref{Phase_diagram_M4}) show that the imbalance can be used to obtain a rough phase diagram with smeared boundaries. However, we also observe as in Ref.~\onlinecite{purkayastha2017nonequilibrium} a non-monotonic decrease in the imbalance even if we cross from a localized regime to a ME regime (one would expect a strict decrease, as the number of localized states decreases). The fact that the steady state imbalance is not a trivial function of the fraction of localized states is also seen in Models 2 and 4. A uniform fraction of localized states should have implied an uniform value for the steady state imbalance, but this not the case in the localized regime of Models 2 and 4, as seen in Fig.~\ref{Phase_diagram_M2}(b) and \ref{Phase_diagram_M4}(b).

\begin{figure}[t]
\begin{tikzpicture}

\node (img9) {\includegraphics[height = 6cm, width=8cm]{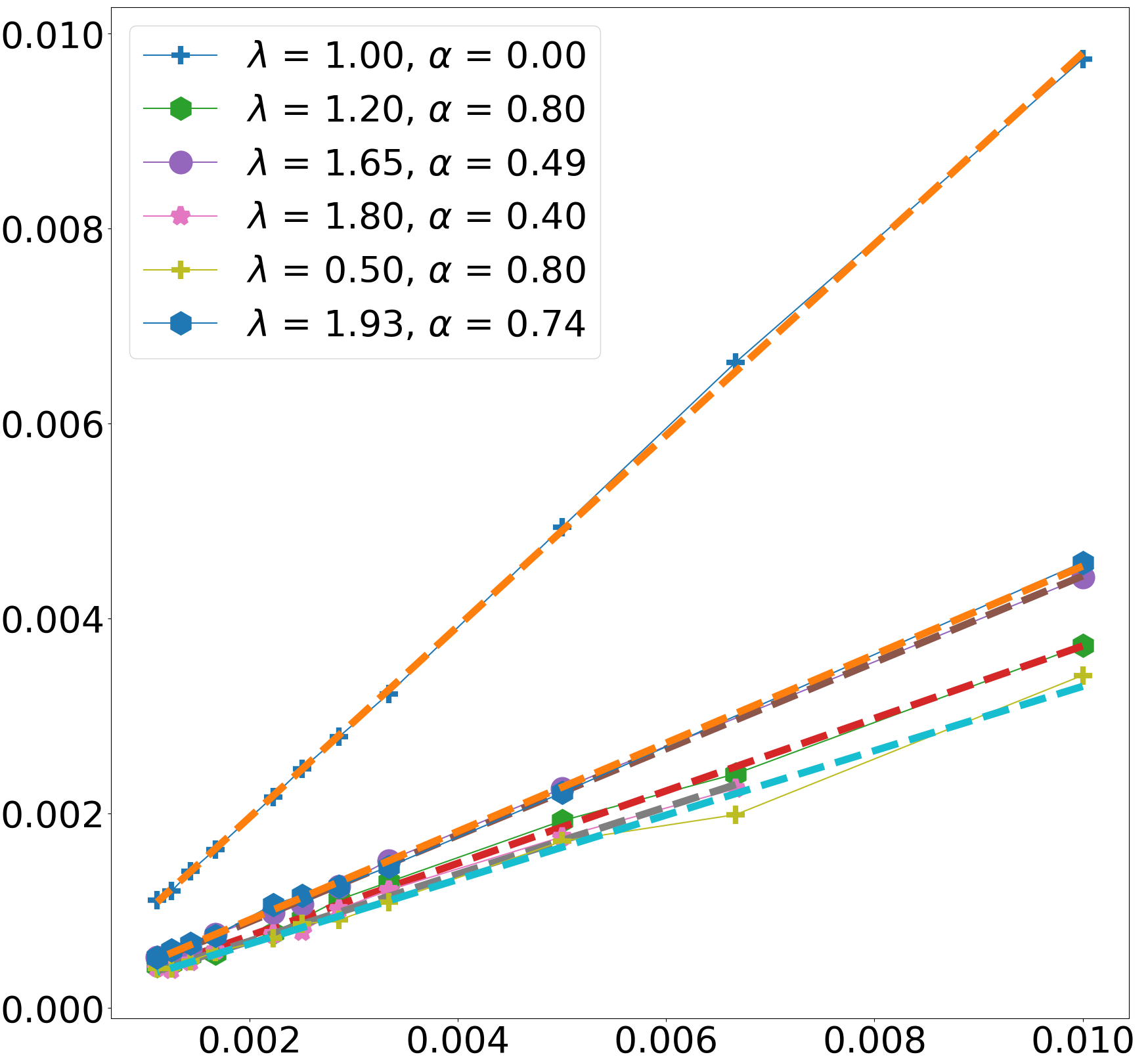}};
\node[left=of img9,node distance=0cm,rotate=90,anchor=center,yshift=-0.6cm]{\large{ $ I-I_{0} $}};
\node [below= of img9,node distance=0.2cm,yshift=1.25cm]{\large{$1/L$}};
\node[below= of img9] (img10) {\includegraphics[height = 6cm, width=8cm]{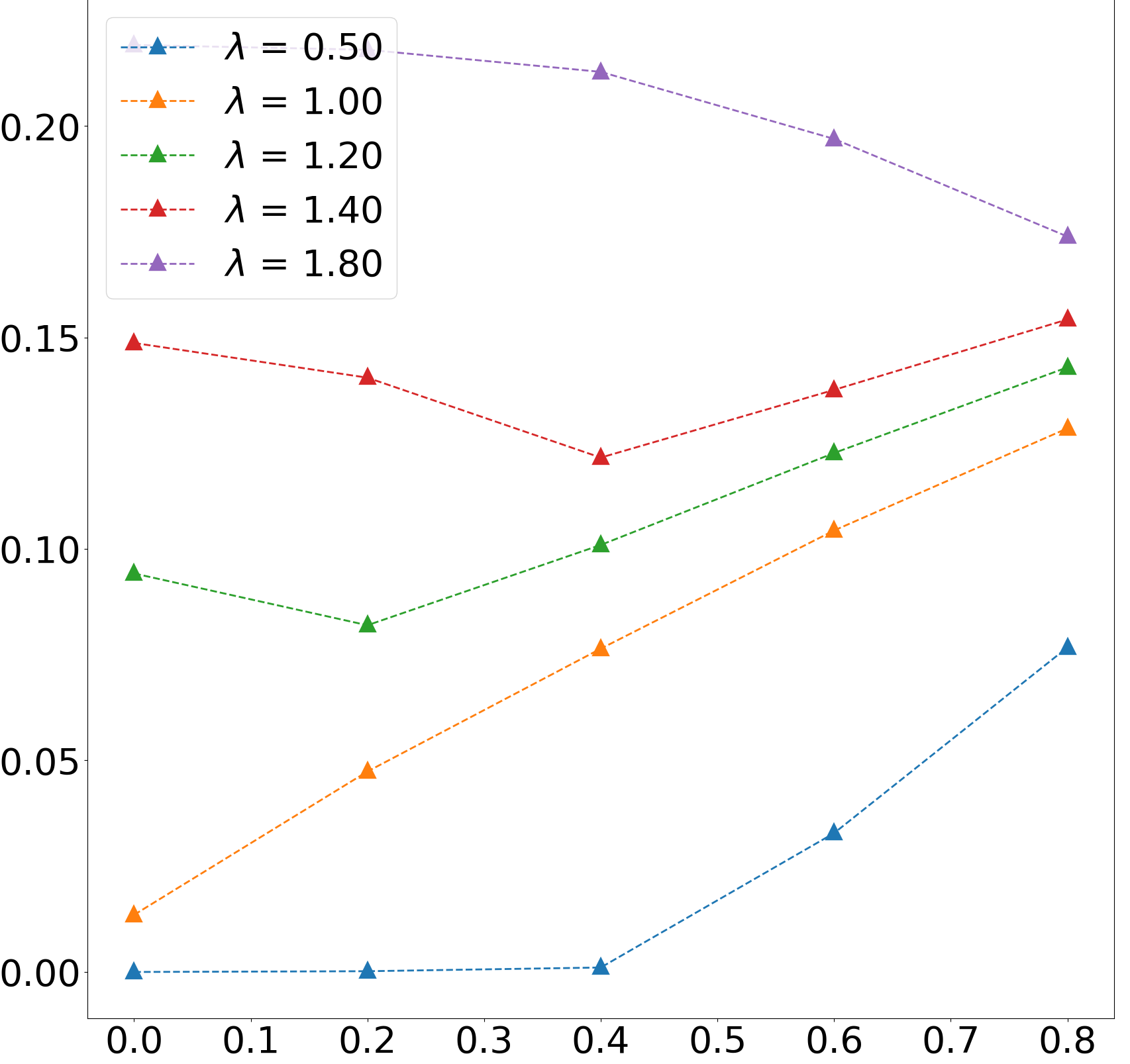}};
\node[left=of img10,node distance=0cm,rotate=90,anchor=center,yshift=-0.6cm]{\large{ $I_0$}};
\node [below= of img10,node distance=0.2cm,yshift=1.25cm]{\large{$\alpha$}};
\end{tikzpicture}
\caption{(Top Panel) Dependence of steady state imbalance on the system size $L$, averaged over 100 realizations for   Model 1. The residual value at thermodynamic limit has been subtracted to make the plots converge at $L \rightarrow \infty$ which makes the 1/L linear scaling clear. 
(Bottom Panel) Thermodynamic limit of steady state imbalance, $I_{0}$, averaged over 100 realizations, as function of $\alpha$ for fixed $\lambda$ for Model 1, obtained from the linear fits for the top panel. Notice that there is a change in trend of $I_{0}$ whenever there is a phase transition from a localized to mobility edge phase, for $\lambda = 1.2$ and $\lambda = 1.4$.} 
\label{I_0}
\end{figure}
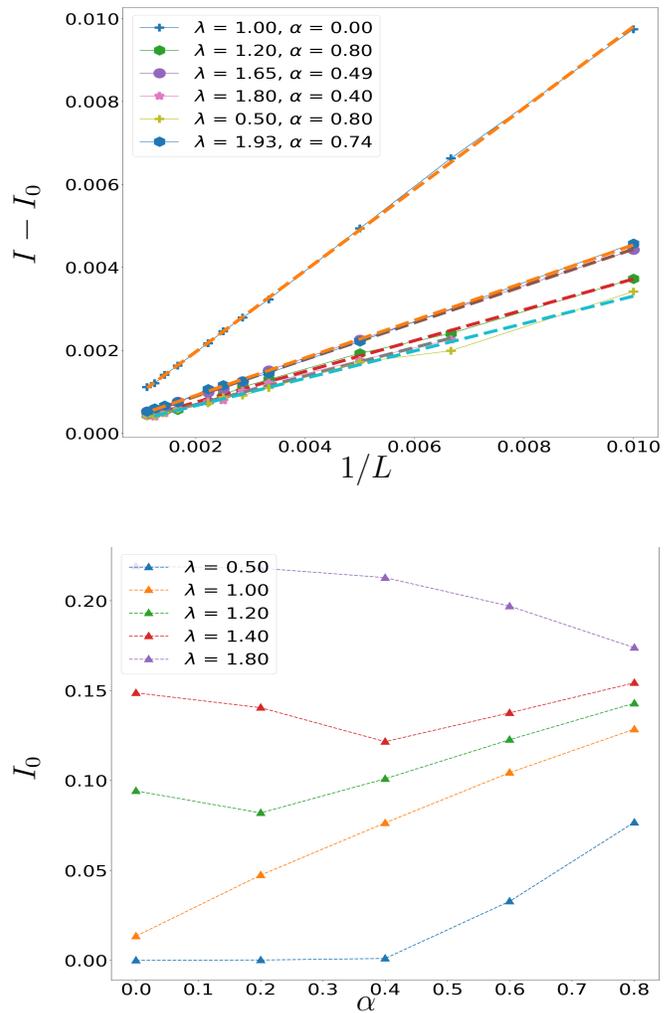

To exclude any finite size effects that may cause this non-monotonic  behaviour of the steady state imbalance, we obtain a scaling of the steady state imbalance with system size. To leading order, the correction appears to be of the form $I = \frac{a}{L} + I_{0}$, presented in Fig.~\ref{I_0}(a). Thus, the thermodynamic limit is the intercept of the straight line, $I_{0}$.

\begin{table*}[t!]
\begin{tabular}{|c|c|c|}
\hline
      Type of spectrum  & Model & Parameters ($\lambda,\alpha$)  \\
     \hline
     Mobility Edge & 1 & (0.5,0.8),(1.2,0.8)   \\
                   & 2 & (0.5,0.8),(0.75,0.8) \\
                   & 3 & (0.8,0.2),(1.40.4)*,(1.7,0.5) \\
                   & 4 & (0.75,0.2), (0.9,0.8)  \\
     \hline
     Localized & 1 & (1.8,0.4) \\
               & 2 & (1.5,0.1),(1.5,0.8) \\
               & 3 & (1.7,1.1),(2.00,0.2)\\
               & 4 & (1.4,0.3),(1.8,0.2)\\
     \hline
     AAH Critical Point & 1,2,3 & (1.0,0)\\
                                & 4   & (1.0,1.0)  \\
     \hline
     Critical Point(ME to Localized transition)&1& (1.65,0.49),(1.93,0.74) \\
     \hline

\end{tabular}

\caption{ Finite size dependence of steady state imbalance. The values shown in the third column of the table are of the representative parameters of the corresponding model identified in the second column. The corresponding type of single particle spectrum is given in the first column. The points marked with $``*"$ are the multiple ME zones (Appendix \ref{Mobilityedges_M3_appendix}). Note that in all cases the steady state imbalance has $1/L$ scaling. For the critical cases (i.e., third and fourth row of the table), this $1/L$ scaling happens even at small system sizes as per our numerics. } 
\label{Finite_size_Table}
\end{table*}
 At the critical point, our numerics show that $1/L$ behaviour survives even for small system sizes, verified for Models 1,2 and 4. Here, the localization length is of the order of system size along the line separating the localized and ME phase.   The $1/L$ scaling also seems to hold at small system sizes at the AAH critical point . The leading order correction seems to hold quite well whenever the imbalance has a large value, even in the localized and ME phases. We emphasize this is only the leading order term, and an analytical expression for this quantity in terms of the parameters of the model is currently lacking.

 The linear fits in Fig.~\ref{I_0}(a) are calculated in various regions of the ME phase, and the localized regime of the model, as summarized in Table \ref{Finite_size_Table} , along with the 3 other models, which we present in the appendix

The non-monotonic behaviour of $I_{0}$ in Fig.~\ref{I_0} points out that the non monotonicity in the steady state imbalance, as seen here and in Ref.~\onlinecite{purkayastha2017nonequilibrium} is not a finite size effect , but rather tells us that the imbalance is no longer a trivial function of the no of localized and extended single particle states.

\begin{figure}[b]
\begin{tikzpicture}

\node (img11) {\includegraphics[height = 6cm, width=7cm]{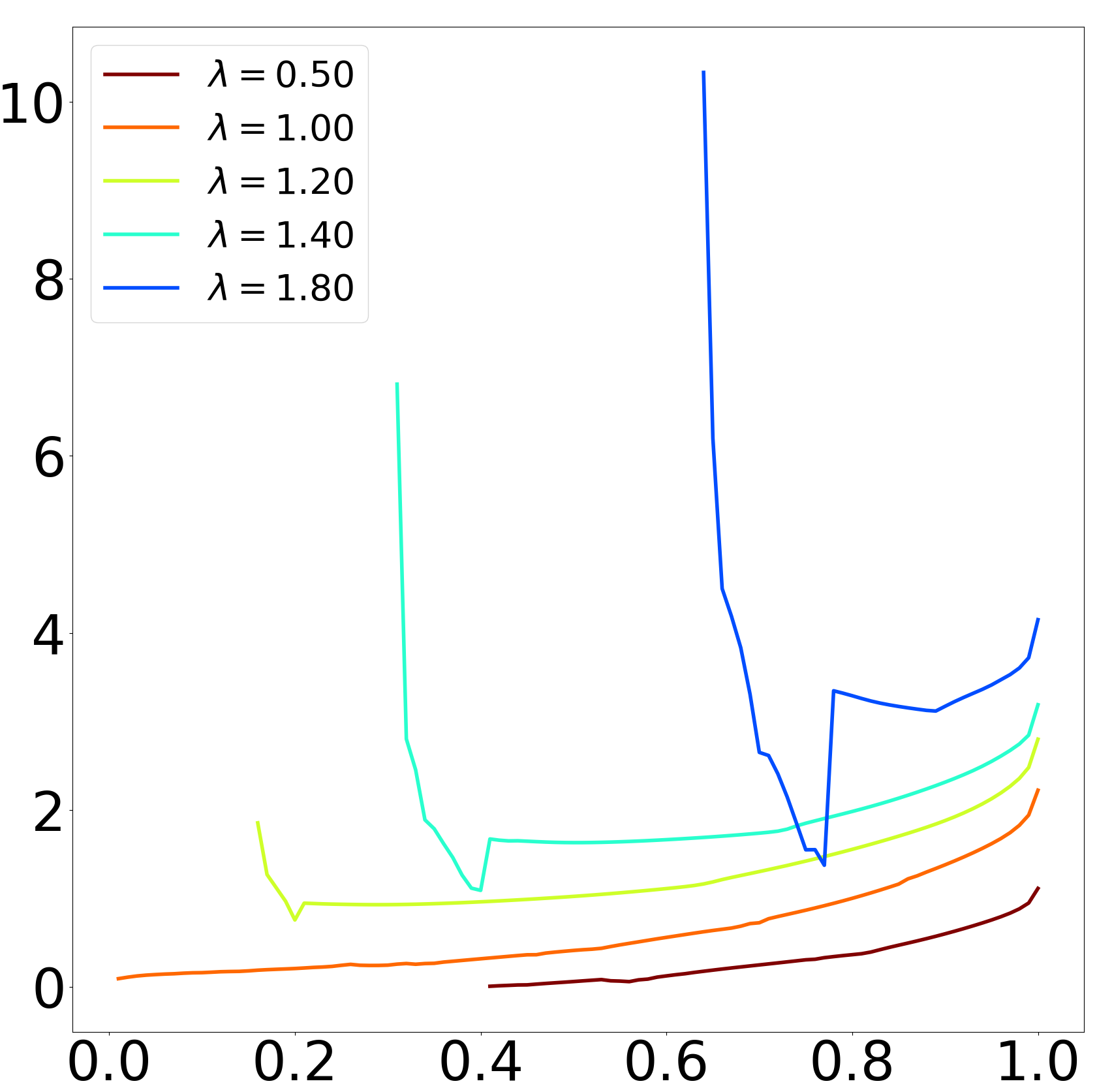}};
\node[left=of img11,node distance=0cm,rotate=90,anchor=center,yshift=-0.6cm]{\large{ $\epsilon$}};

\node (img12)[below= of img11,yshift = 0.85cm] {\includegraphics[height = 6cm,width=7cm]{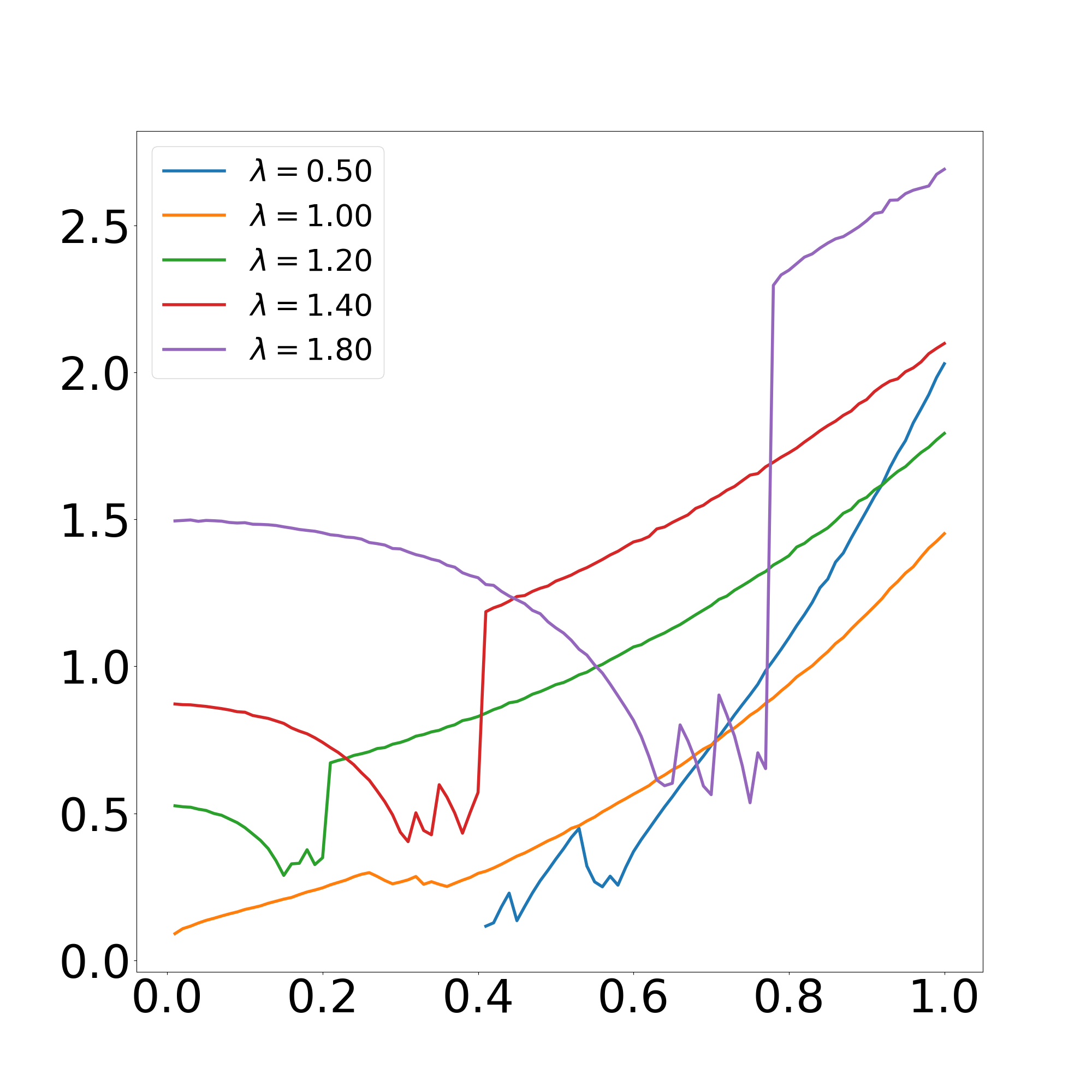}};
\node[left=of img12,node distance=0cm,rotate=90,anchor=center,yshift=-0.6cm]{\large{$\epsilon'$}};
\node [below= of img12,node distance=0.2cm,yshift=1.25cm]{\large{$\alpha$}};
\end{tikzpicture}
\caption{
(a) $\epsilon$  (Eq.~\ref{epsilon}) for Model 1, for $L= 256$, averaged over 100 realizations. $\epsilon$ is calculated only in the mobility edge phase, along a constant $\lambda$ line in the phase diagram. (b) $\epsilon'$ (Eq.~\ref{epsilonp}) for different values of $\lambda$. In the localized phase, it is calculated over the entire spectrum, whereas in the mobility edge phase it is calculated only over states below the mobility edge that are localized. Note that both $\epsilon$ and $\epsilon'$ are 0 in the delocalized phase.}
\label{Epsilon}
\end{figure}
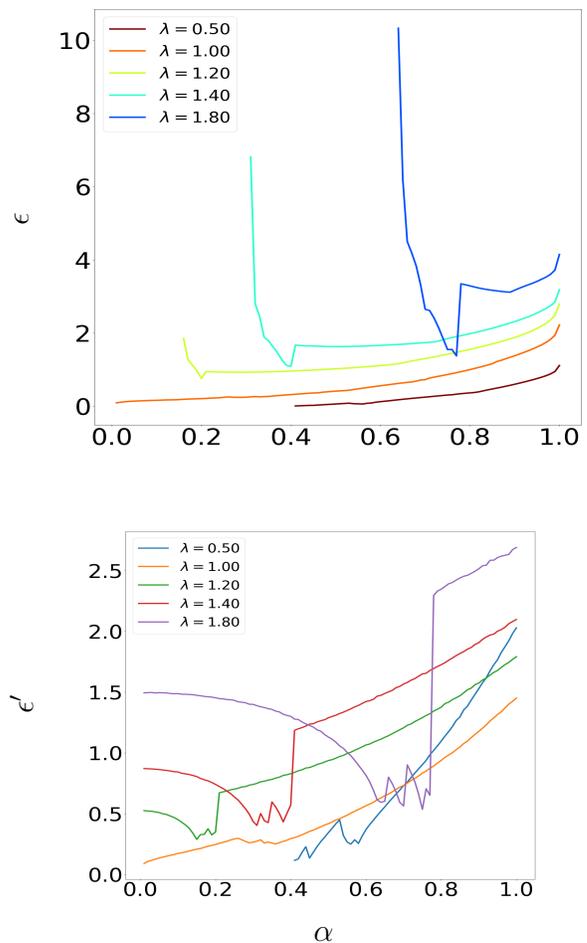

\subsection{Dimensionless parameters $\epsilon$ and $\epsilon'$}\label{Ep}

 The anomalies in steady state imbalance in Ref.~ \onlinecite{purkayastha2017nonequilibrium} and in Fig.~\ref{I_0}  show that it is not a trivial function of the fraction of localized states. The degree to which an initial CDW can relax depends on the single particle states the particles occupy, and hence the localization length of theses states should also dictate how far the particles can travel and how much the CDW can relax. A useful quantity is the dimensionless parameter $\epsilon$ defined in
Ref.~\onlinecite{modak2018criterion}, which calculates how
strongly the single particle localized states are localized versus how strongly the single particle extended states are extended. The definition of this parameter was presented in Section III, along with the new parameter, $\epsilon'$ for calculating the strength/degree  of localization of the single particle localized states. The plots for the two parameters are presented in Fig.~\ref{Epsilon}. 

The plot for $\epsilon'$ shows that even within the localized phase, not all the states are uniformly localized, some are localized more than the others, which is expected. But it is the presence of this non uniformity in the localization length of these states that gives rise to the non uniformity of the steady state imbalance in the localized phase. The trend of $I_{0}$ in Fig.~\ref{I_0} in the localized phase is consistent with that of $\epsilon'$. This is also the case for the other three models, presented in the appendix.

In case of the mobility edge phase, there is a competition between the extended and localized states. The extended states allow the particle to move through the entire length of the lattice and hence allow the CDW to relax completely, whereas the localized states restrict the motion of the particles and creates a configuration that is closer to the initial condition. Thus, the $\epsilon$ parameter measures how effective this competition is by taking into account not only how many of these extended/localized states are present but also how much each state contributes to the CDW relaxation.

A direct tallying of  Fig.~\ref{I_0}(b) and Fig.~\ref{Epsilon} shows that this is indeed the case. In the mobility edge phase, $I_{0}$ follows the trend of the parameter $\epsilon$, it increases in accordance with $\epsilon$ for $\lambda = $ 0.5,1,1.2,1.4, and decreases in accordance with $\epsilon$ for $\lambda$ = 1.8.

\begin{figure}[H]
\begin{tikzpicture}
\node (img11) {\includegraphics[height = 6cm, width=7cm]{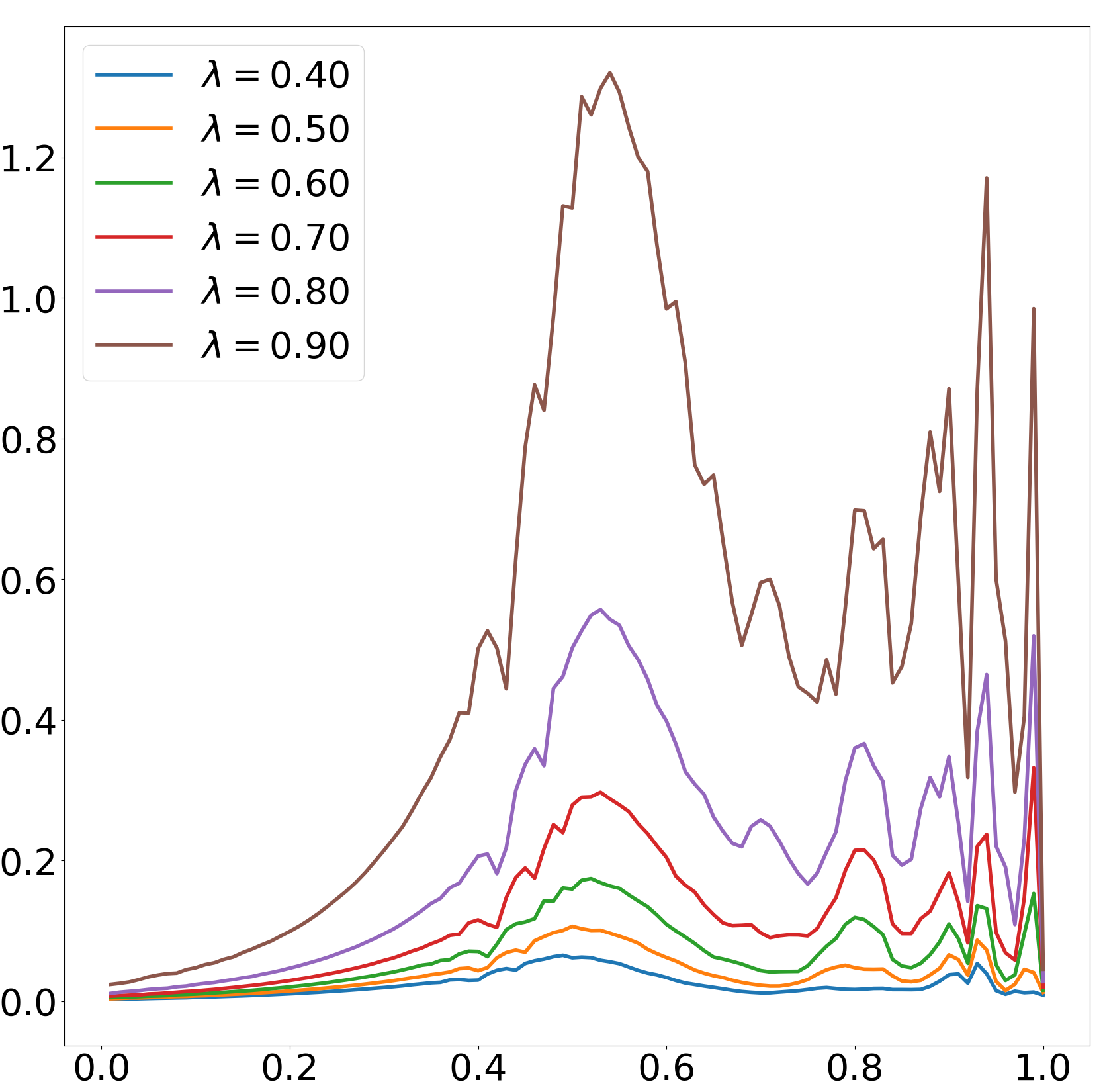}};
\node[left=of img11,node distance=0cm,rotate=90,anchor=center,yshift=-0.6cm]{\large{ $\epsilon$}};

\node (img12)[below= of img11,yshift = 0.85cm] {\includegraphics[height = 6cm,width=7cm]{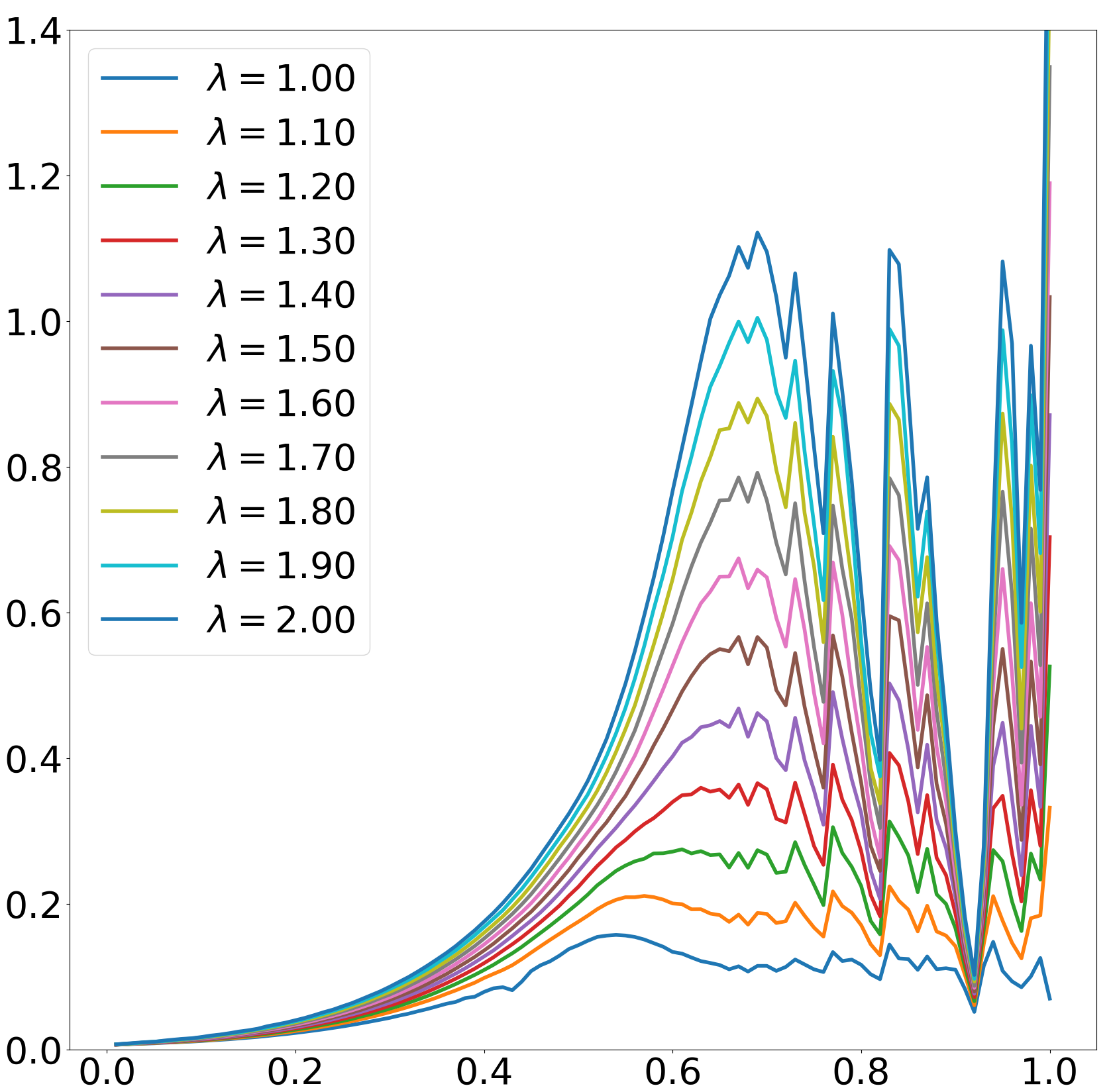}};
\node[left=of img12,node distance=0cm,rotate=90,anchor=center,yshift=-0.6cm]{\large{$\epsilon'$}};
\node [below= of img12,node distance=0.2cm,yshift=1.25cm]{\large{$\alpha$}};
\end{tikzpicture}
\caption{
(a) $\epsilon$  for Model 4, for $L= 256$, averaged over 100 realizations. Note the relative peaks in $\epsilon$ around $\alpha = 0.5$, in the mobility edge phase, although the imbalance phase diagram is structure-less in this region and doesn't show any region of relatively high imbalance. (b) $\epsilon'$ for Model 4 averaged over 100 disorder realizations, calculated for different values of $\lambda$ in the localized phase. Note the dip in the value for $\epsilon'$  for all $\lambda$'s around $\alpha = 0.9 $. This corresponds to a finger like projection of low imbalance region into the localized phase [Fig.~\ref{Phase_diagram_M4}(b)].} 
\label{Epsilon_M4}
\end{figure}
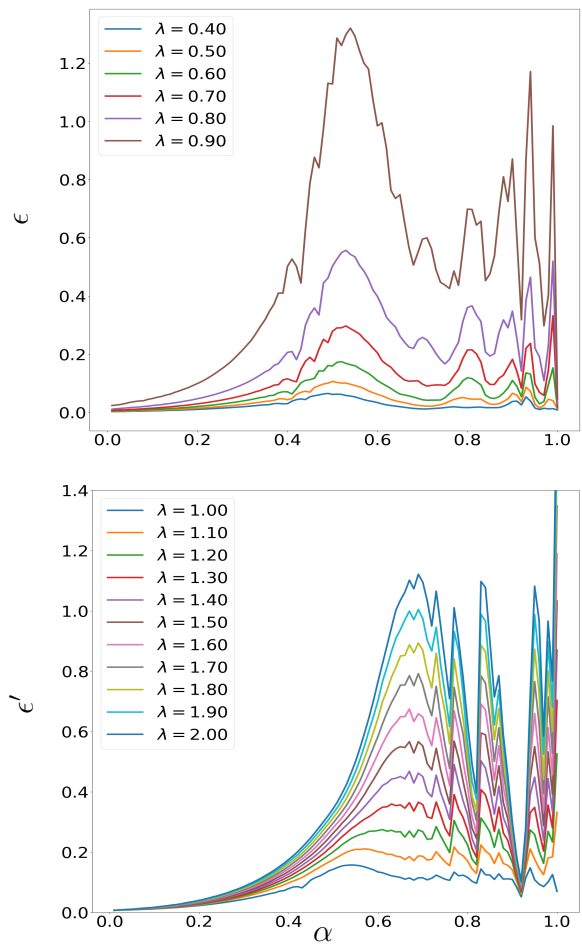

One key thing to notice from Fig.~\ref{I_0} is that the behaviour of $I_{0}$ is also dependent on the phase the model is in. In other words, it increases or decreases differently in the ME phase and the localized phase . Also, a change in the nature of $I_{0}$ as we increase $\alpha$ keeping $\lambda$ fixed indicates that there has been a phase transition, as expected from the imbalance diagram in Fig.~\ref{Phase_diagram_M1} - \ref{Phase_diagram_M4}. 
 A more convincing point for relevance of these two parameters to explain the steady state imbalance is made by the imbalance phase diagram of Model 4. In Model 4 (Fig.~\ref{Phase_diagram_M4} ), there is a finger like projection of a low  imbalance region into the localized phase, and the imbalance is not constant across $\lambda > 1$, although every $(\lambda,\alpha)$ for $\lambda > 1$ corresponds to a localized phase. Plots of $\epsilon$ and $\epsilon'$ in Fig.~\ref{Epsilon_M4} shows how this might happen. The $\epsilon'$ is very low for all $\lambda$ near $\alpha = 0$. The eigenstates are not very strongly localized, they have very large localization length, of order $N$ (system size) and hence the imbalance is close to $0$. The localized character of these states start getting stronger and stronger, and the parameters $\epsilon'$ peaks with this, reaching it's maximum in between $\alpha = 0.7$ and $\alpha = 0.8$. This is followed by a dip in $\epsilon'$ around $\alpha = 0.9$, where the states seem to completely lose their localized form, and this is the region of low imbalance that is seen in Fig.~\ref{Phase_diagram_M4}(b). The states pick up their localization strength soon after, and the collective localization strength reaches its maximum when $\alpha =1$, and the system hits the AAH limit.

 Table II contains a brief summary of the finite size scaling explored in this paper for different models. At the critical point, we find from numerics that $1/L$ scaling holds even at small system sizes, and in the localized and mobility edge phase, this is the leading order correction term. Plots for the other three models are in given in the appendix. 
\vspace{0.05cm}

\section{Summary}\label{conclusion}

To summarize, employing numerical exact diagonalization on systems of  length up to 900 sites,  we have seen that we may define a metal to insulator transition in  non-interacting (interacting) one-dimensional quasi periodic systems with the population imbalance as an order parameter. A value close to zero indicates a metallic (ergodic) phase, whereas a value close to the initial starting point defines an insulating (MBL) phase. An intermediate value of the imbalance indicates restrictive relaxation of the charge density wave, and hence presence of both localized and delocalized states (i.e., a mobility edge phase). 

The imbalance has a linear dependence on inverse length to leading order for all the systems studied here, in the localized and mobility edge phase, i.e. wherever the imbalance value was high enough to allow us to determine such a dependence. We see that while this $1/L$ behaviour holds even at small system sizes at the critical point, it is also the leading order correction term for all the systems considered above. We have extracted the thermodynamic behaviour of imbalance from this finite size scaling.

The imbalance shows non-monotonic behaviour, as found in the phase diagram of imbalance for Model 1 in Ref.~\onlinecite{purkayastha2017nonequilibrium}. We see that although we cross from a region of partial localization of the single particle spectrum to one with complete localization, the imbalance drops, around $\lambda$ = 1.2. Model 2 (in appendix) illustrates this in a better way, where we see non-monotonic behaviour even in the localized phase, where one might naively have expected the imbalance to stay same, as all the states are localized (Fig.~\ref{I_0_appendix}). This indicates that the fraction of localized states is not the only factor deciding the relaxation of the CDW, but also how effectively the localized states are localized, as this decides the length over which the particles may move.

We calculate the, $\epsilon$\cite{modak2018criterion}, which is a quantifier of how effectively localized the single particle localized states are versus how strongly delocalized the single particle delocalized states are. A larger value implies a larger influence of the localized states, and hence a higher value of imbalance. The thermodynamic limit of the imbalance $I_{0}$ follows this parameter in the mobility edge phase for all the systems we considered, except for Model 3, where the calculation is unfeasible due to an absence of analytical formula for the critical point/mobility edge.

The definition of $\epsilon$ renders it infinity in the localized phase, and hence , if one is interested in the localized regime, a different parameter $\epsilon'$ may be used, which tells us how strongly localized the single particle localized states are. This quantifies the degree of movement of the particles in the localized phase, and explains the non-monotonicity, like the one in Fig.~\ref{I_0_appendix}(c) in the localized phase. The imbalance follows this parameter in the localized phase for all the systems we considered, except for Model 3, where the calculation is unfeasible. to conclude, the parameters $\epsilon$ and $\epsilon'$ play a pivotal role in understanding the phase diagrams of our models.

 As a future outlook, extending these studies to the interacting case will be very interesting both from a theoretical and an experimental perspective. Rigorous theoretical understanding of the $1/L$ behaviour of the steady state imbalance remains an open question. Studying higher spacial dimensional systems \cite{devakul2017anderson} through the lens of the diagnostics we introduced is an important future goal. 

\section{Acknowledgements}

We would like to thank R. Modak, A. Purkayastha, S. Ghosh,  F. Weiner, F. Evers, A. Acharya and S. Sarangi  for useful discussions. MK would like to acknowledge support from the project 6004-1 of the Indo-French Centre for the Promotion of Advanced Research (IFCPAR), 
Ramanujan Fellowship (SB/S2/RJN-114/2016), 
SERB Early Career Research Award (ECR/2018/002085) and SERB Matrics Grant (MTR/2019/001101) from the Science
and Engineering Research Board (SERB), Department of
Science and Technology (DST), Government of India. SM thanks Quantum Information Science and Technology (QuST) initiative of the DST, Government of India. We are grateful to the ICTS-TIFR high performance computing facility.

\appendix

\section{System size scaling for other models}\label{FSS_appendix}

The system size scaling obtained for Model 1 is quite robust and holds generally for all other models we consider in this paper (Table~\ref{Models}) . Fig.~\ref{I_0_appendix} (a) shows that even for Model 2 in the strict localized phase, the imbalance does change from one point to the other,  although the number of localized states stay the same. This further supports the hypothesis that the number of localized states is not the only deciding factor for the steady state value of imbalance, but also how effectively they are localized against the delocalized states (refer to Section IV B and C for the anomalies in Model 1 and subsequent justification). The steady state value of imbalance suggests that the localized states close to $\lambda = 1$ are weakly localized, and they get more and more localized as the system moves deeper into the localized phase. 

Fig.~\ref{I_0_appendix}(b) tells a similar story to the phase diagram Fig.~\ref{Phase_diagram_M3} for Model 3, there is a tiny region of high imbalance around $\alpha = 0.2$ for $\lambda < 1$. For $\lambda > 1$, the imbalance decreases from the localized regime to the mobility edge regime.

\begin{figure}[H]
 
    \begin{tikzpicture}
\node (img13) {\includegraphics[width=7cm, height = 6cm]{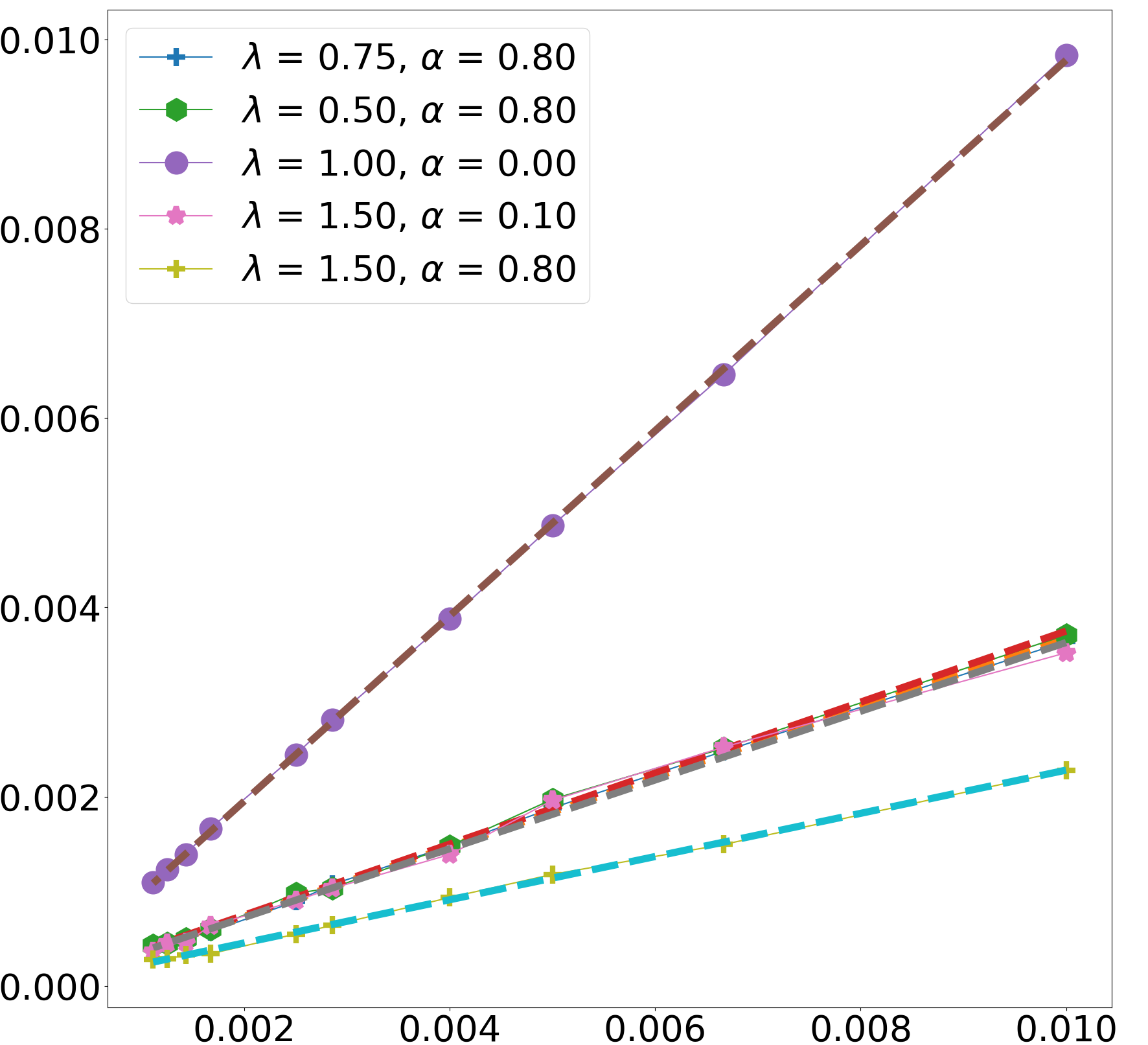}};
\node[left=of img13,node distance=0cm,rotate=90,anchor=center,yshift=-0.80cm]{\large{ $I-I_{0}$}};
\node[left=of img13,node distance=0cm,yshift=2.5cm,xshift=5.8cm]{\large{a}};
    \end{tikzpicture}
 
\begin{tikzpicture}
\node (img14)[below=of img13,xshift = -0.5cm] {\includegraphics[width=7cm, height = 6cm]{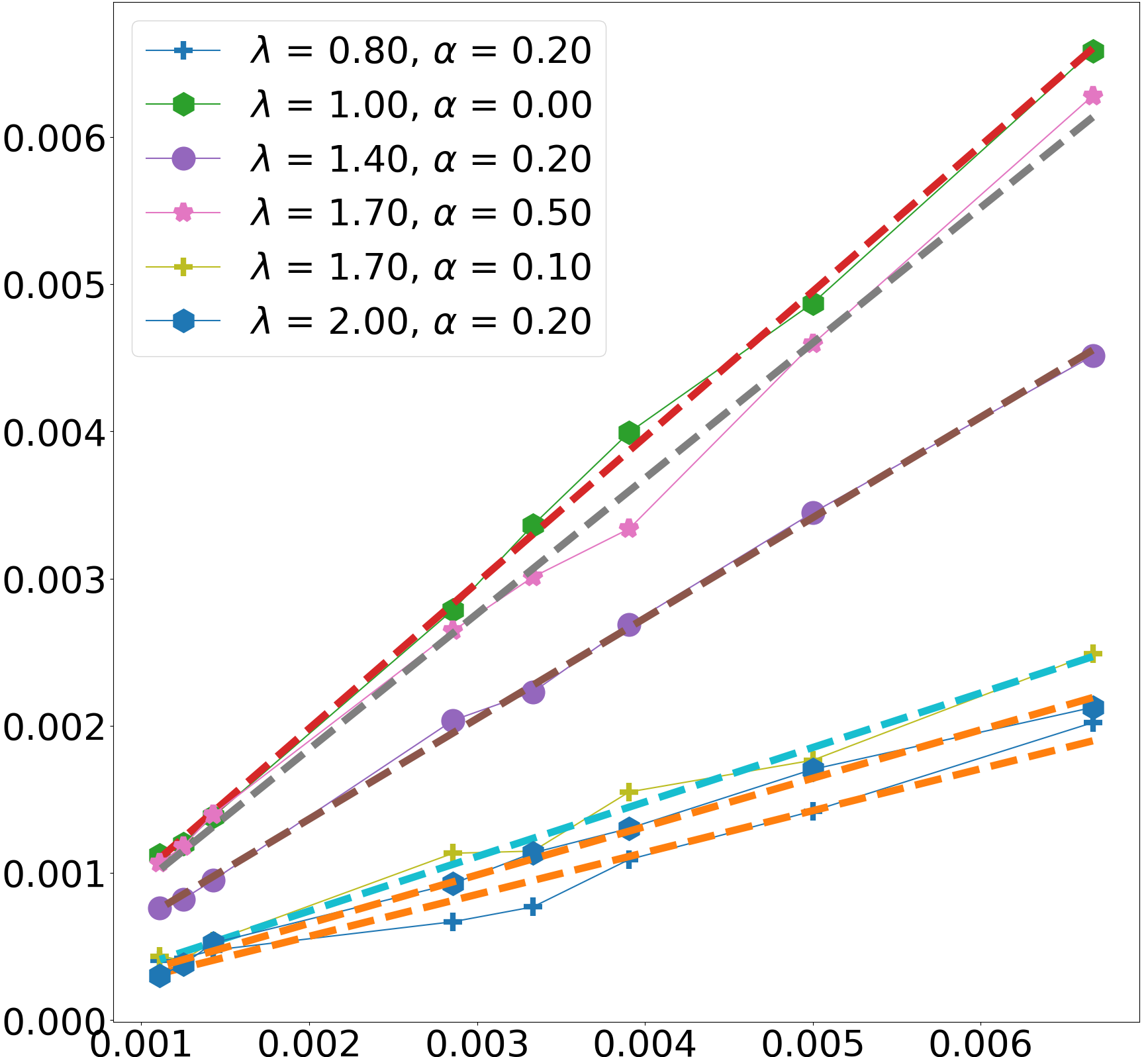}};
\node[left=of img14,node distance=0cm,rotate=90,anchor=center,yshift=-0.8cm]{\large{ $I-I_{0}$}};
\node[left=of img14,node distance=0cm,yshift=2.5cm,xshift=5.8cm]{\large{b}};
\end{tikzpicture}
 
\begin{tikzpicture}
\node (img15)[below= of img14, xshift = 10.2] {\includegraphics[width=7cm , height = 6cm]{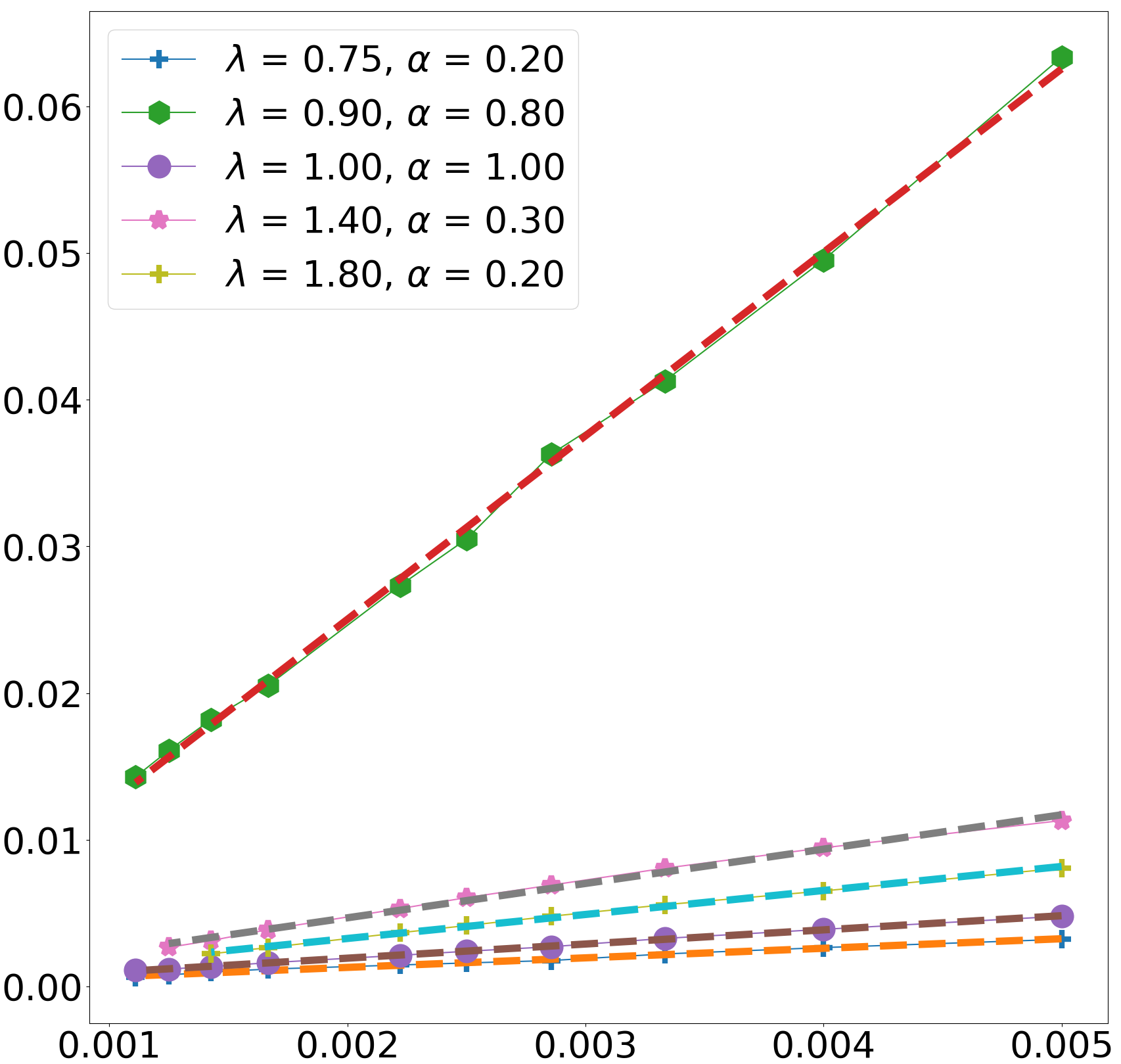}};
\node[left=of img15,node distance=0cm,rotate=90,anchor=center,yshift=-0.8cm]{\large{ $I-I_{0}$}};
\node[left=of img15,node distance=0cm,yshift=2.5cm,xshift=5.8cm]{\large{c}};
\node[left=of img15,node distance=0cm,yshift=-3.5cm,xshift=5cm]{\large{1/L}};
\end{tikzpicture}
 
\caption{a) System size dependence of steady state imbalance, averaged over $100$ disorder realizations for  (a) Model 2 (b) Model 3 (c) Model 4. The residual value at thermodynamic limit has been subtracted to make the plots converge at $L \rightarrow \infty$ and to make the 1/L linear scaling clear.}
    
    \label{Finite_size_appendix}
\end{figure}
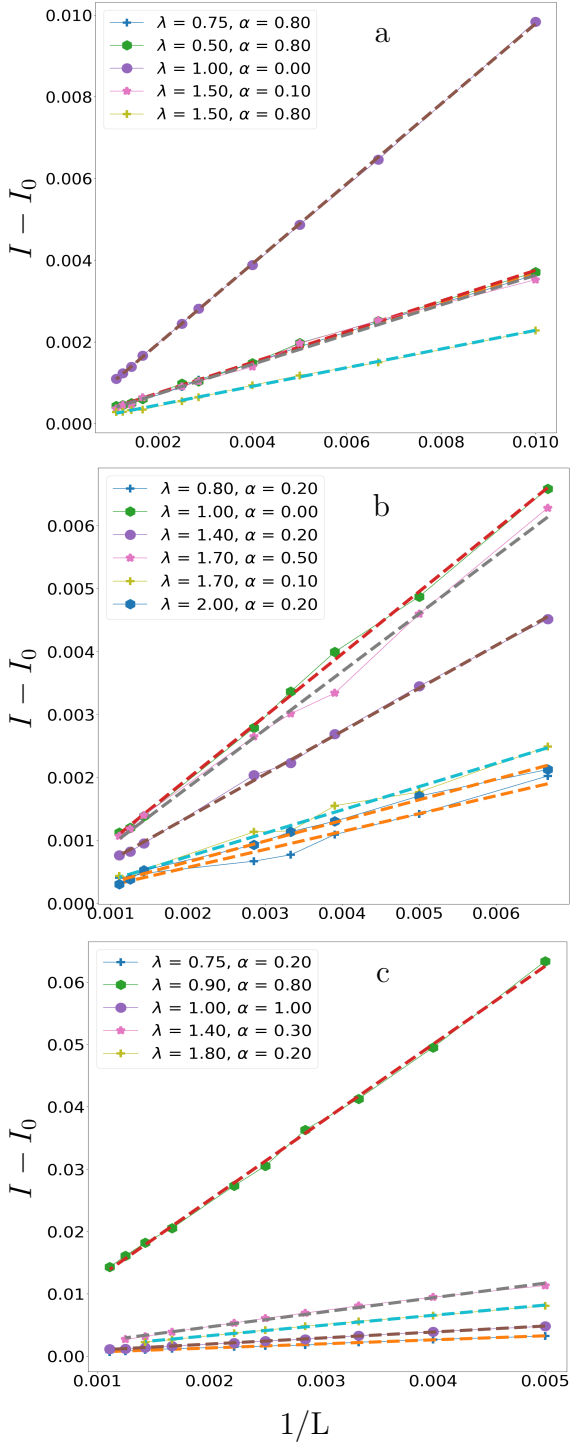

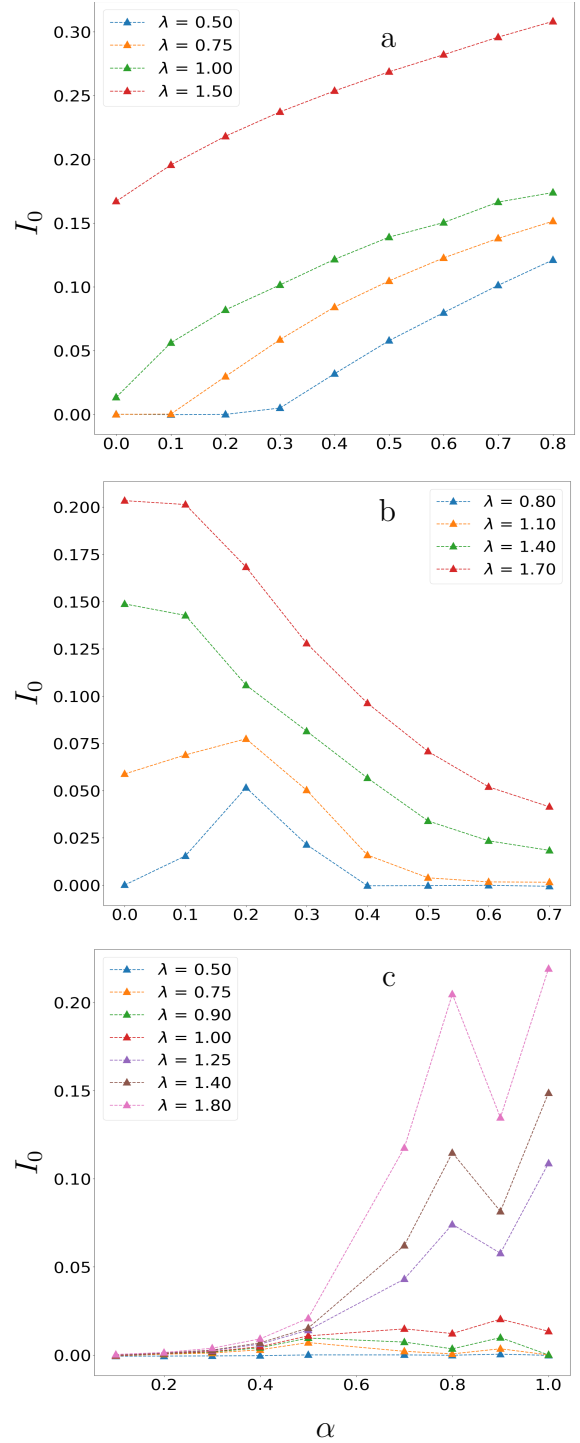
\begin{figure}[H]

    \begin{tikzpicture}
\node (img16) {\includegraphics[width=7cm, height = 6cm]{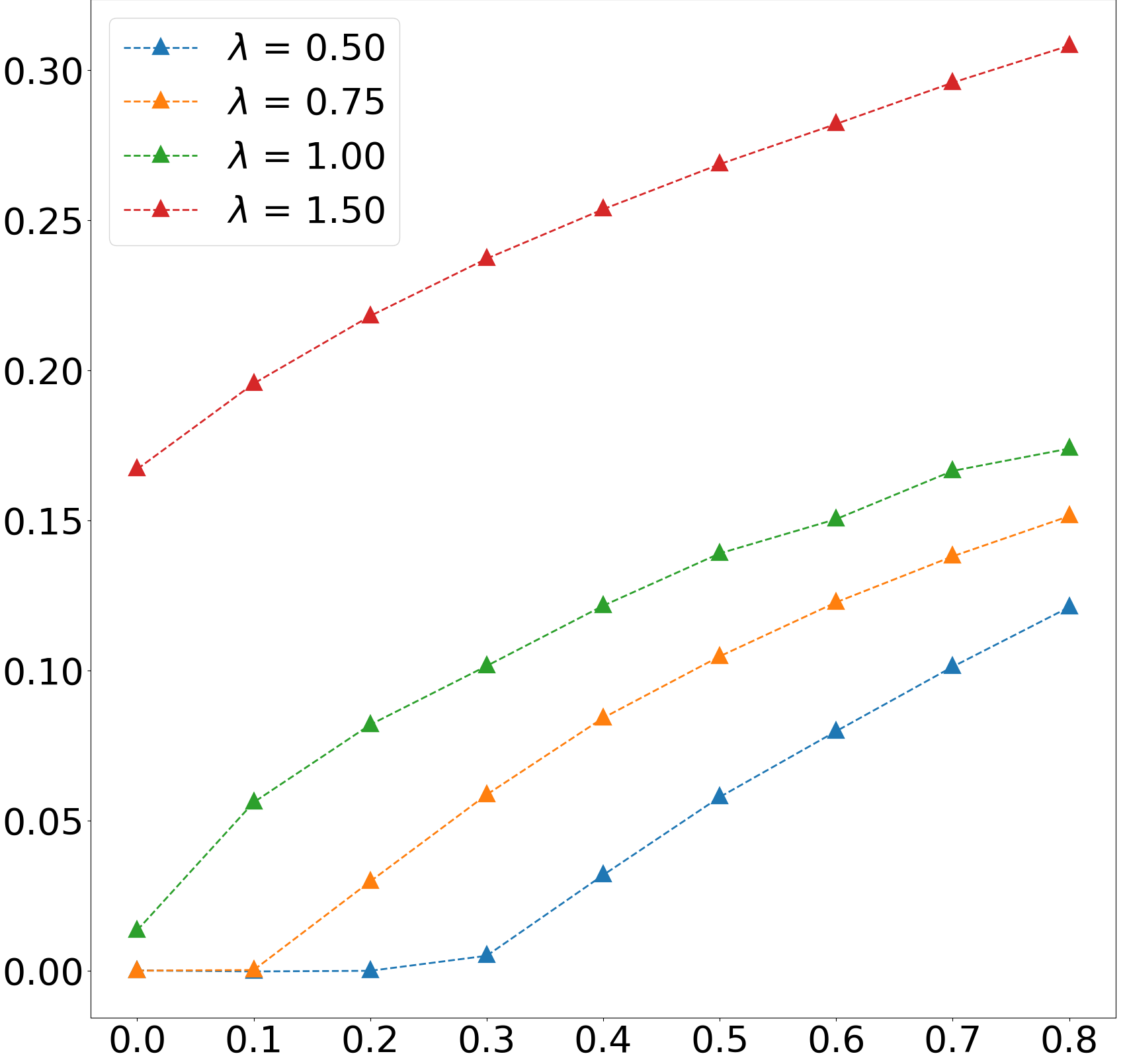}};
\node[left=of img16,node distance=0cm,rotate=90,anchor=center,yshift=-0.80cm]{\large{ $I_{0}$}};
\node[left=of img16,node distance=0cm,yshift=2.5cm,xshift=5.8cm]{\large{a}};
\end{tikzpicture}

\begin{tikzpicture}
\node (img17)[right=of img16,xshift = 0cm] {\includegraphics[width=7cm, height = 6cm]{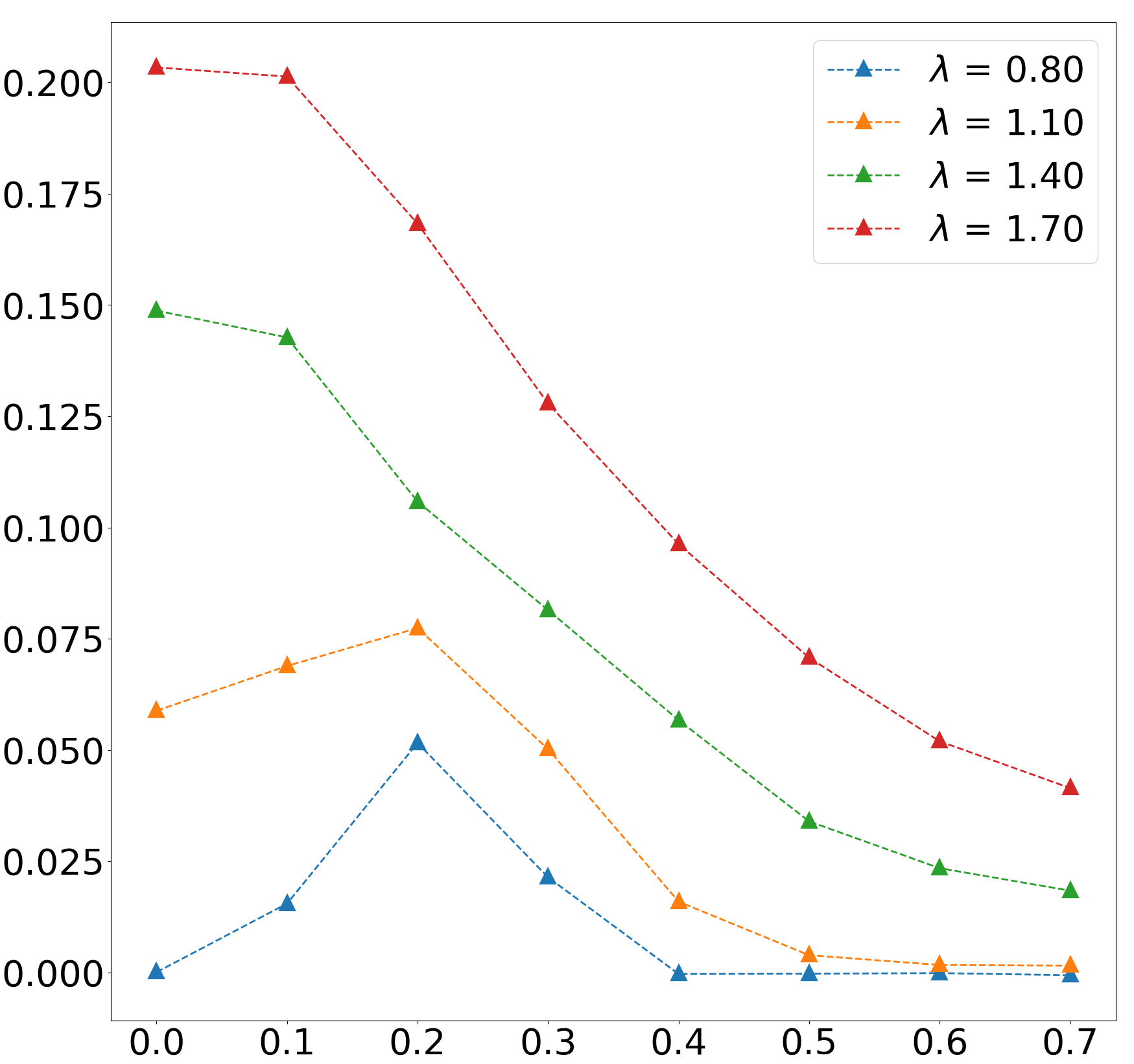}};
\node[left=of img17,node distance=0cm,rotate=90,anchor=center,yshift=-0.8cm]{\large{ $I_{0}$}};
\node[left=of img17,node distance=0cm,yshift=2.5cm,xshift=5.8cm]{\large{b}};
\end{tikzpicture}

\begin{tikzpicture}
\node (img18)[right= of img17, xshift = 0cm] {\includegraphics[width=7cm, height= 6cm]{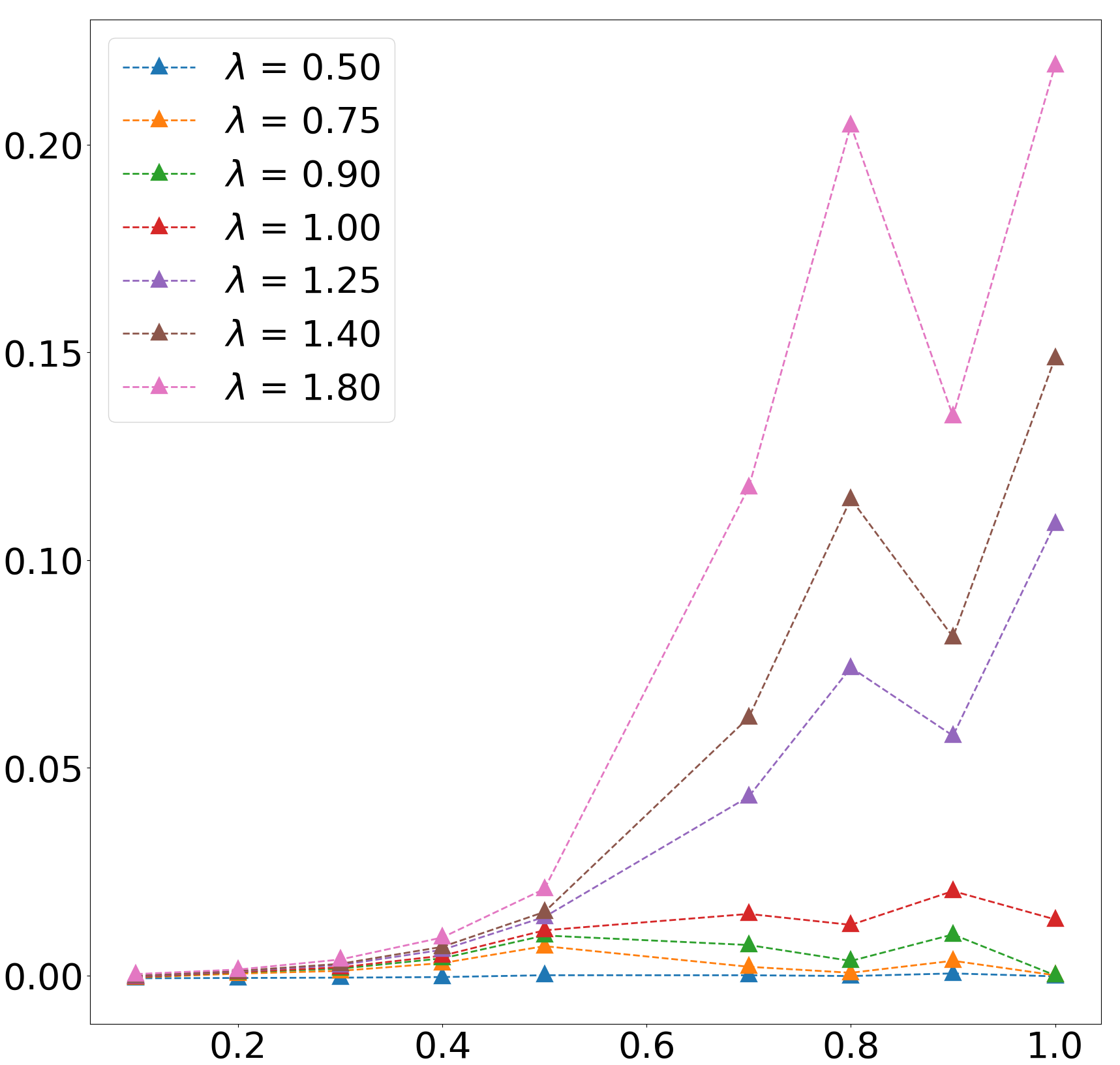}};
\node[left=of img18,node distance=0cm,rotate=90,anchor=center,yshift=-0.8cm]{\large{ $I_{0}$}};
\node[left=of img18,node distance=0cm,yshift=2.5cm,xshift=5.8cm]{\large{c}};
 \node[left=of img18,node distance=0cm,yshift=-3.5cm,xshift=5cm]{\large{$\alpha$}};
\end{tikzpicture}

\caption{ Thermodynamic limit of steady state imbalance, $I_{0}$ for different $\lambda$ with increasing $\alpha$, from similar linear fits (averaged over 100 disorder realizations) up to leading order in $1/L$ like in Fig.~\ref{Finite_size_appendix}. All the linear fits from which we extract $I_0$ have not been presented in Fig.~\ref{Finite_size_appendix} for sake of brevity. Plots are for (a) Model 2,(b) Model 3, (c) Model 4. The various regions explored in Fig.~\ref{Finite_size_appendix} and \ref{I_0_appendix} can be looked up in Table \ref{Finite_size_Table} and the phase diagrams Figs.~ \ref{Phase_diagram_M2},\ref{Phase_diagram_M3},\ref{Phase_diagram_M4}.} 
\label{I_0_appendix}
\end{figure}

 While the AAH critical point has $1/L$ scaling even for small system sizes , for three remaining models (refer to Table \ref{Finite_size_Table}), the localized and mobility edge phases also show a leading order $1/L$ scaling (Fig.~\ref{Finite_size_appendix}). For Model 4, the value of imbalance is pretty low even in the localized phase for all system sizes, and hence the thermodynamic limit of imbalance asymptotes to a value close to zero , except for a tiny sliver of high Imbalance close to $\alpha$ = 0.8 (Fig.~\ref{Phase_diagram_M4}). 

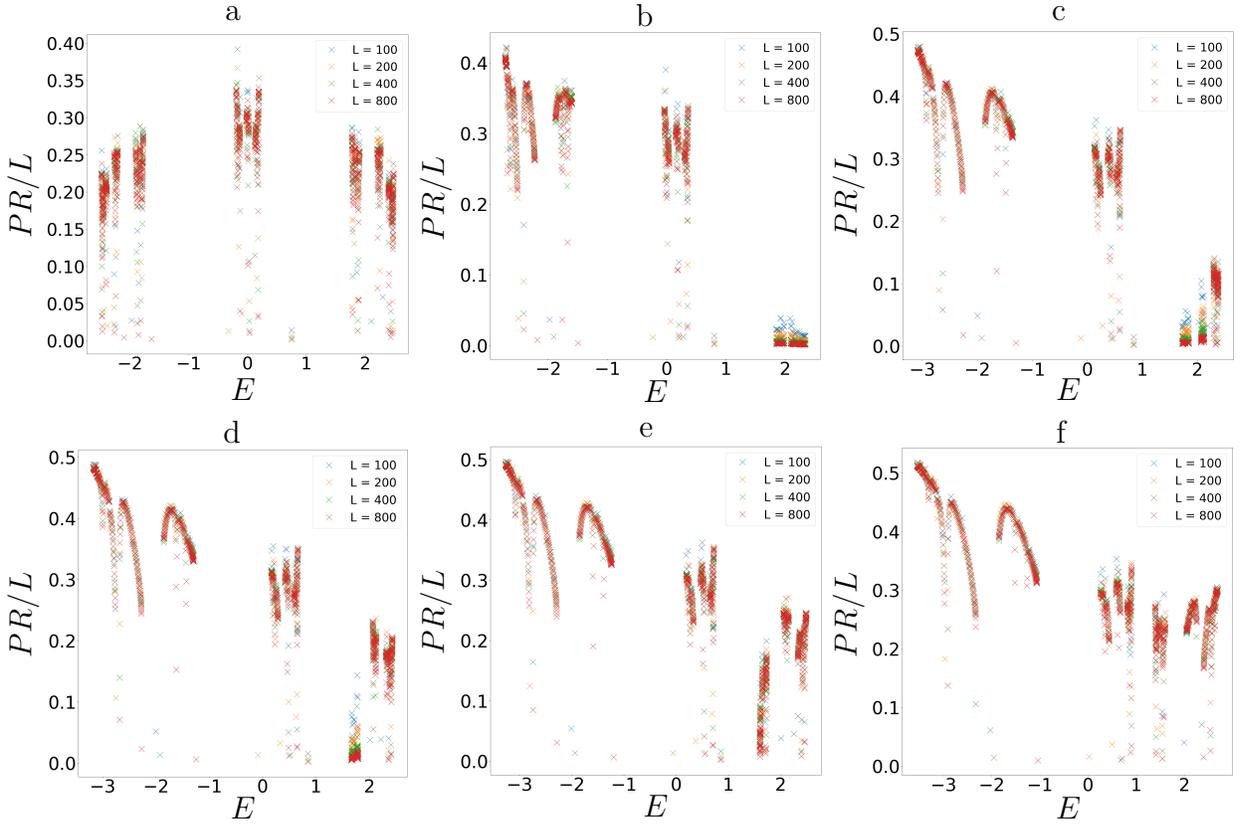
\begin{figure*}[t!]
\centering
\begin{minipage}{0.3\textwidth}
    \begin{tikzpicture}
\node (img22) {\includegraphics[width=0.9\textwidth]{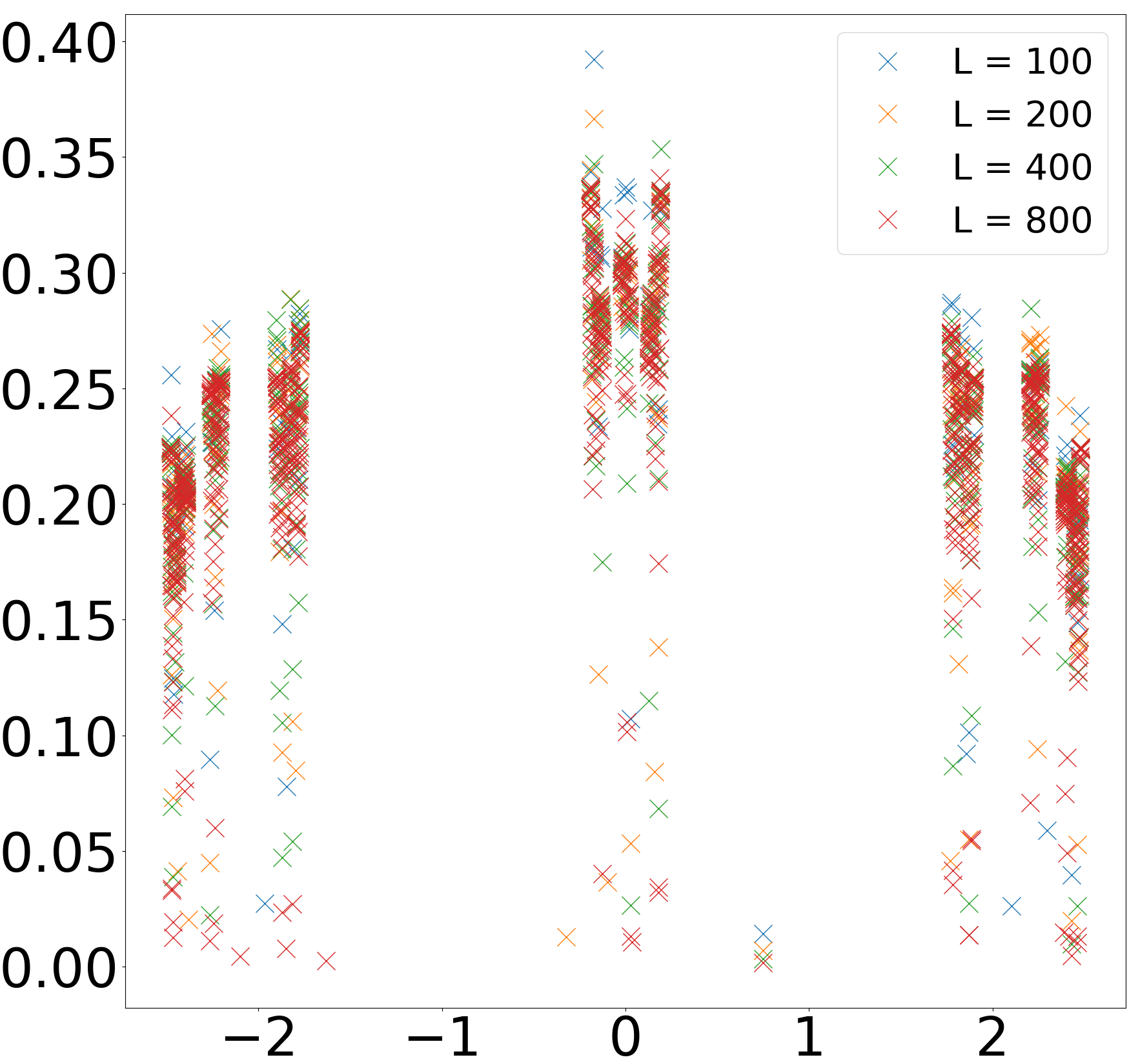}};
\node[left=of img22,node distance=0cm,rotate=90,anchor=center,yshift=-0.80cm]{\large{ $PR/L$}};
\node[left=of img22,node distance=0cm,yshift=2.5cm,xshift=3.8cm]{\large{a}};
\node[left=of img22,node distance=0cm,yshift=-2.5cm,xshift=4.0cm]{\large{$E$}};
\end{tikzpicture}
\end{minipage}
\begin{minipage}{0.3\textwidth}
\begin{tikzpicture}
\node (img23)[right=of img22,xshift = -0.5cm] {\includegraphics[width=0.9\textwidth]{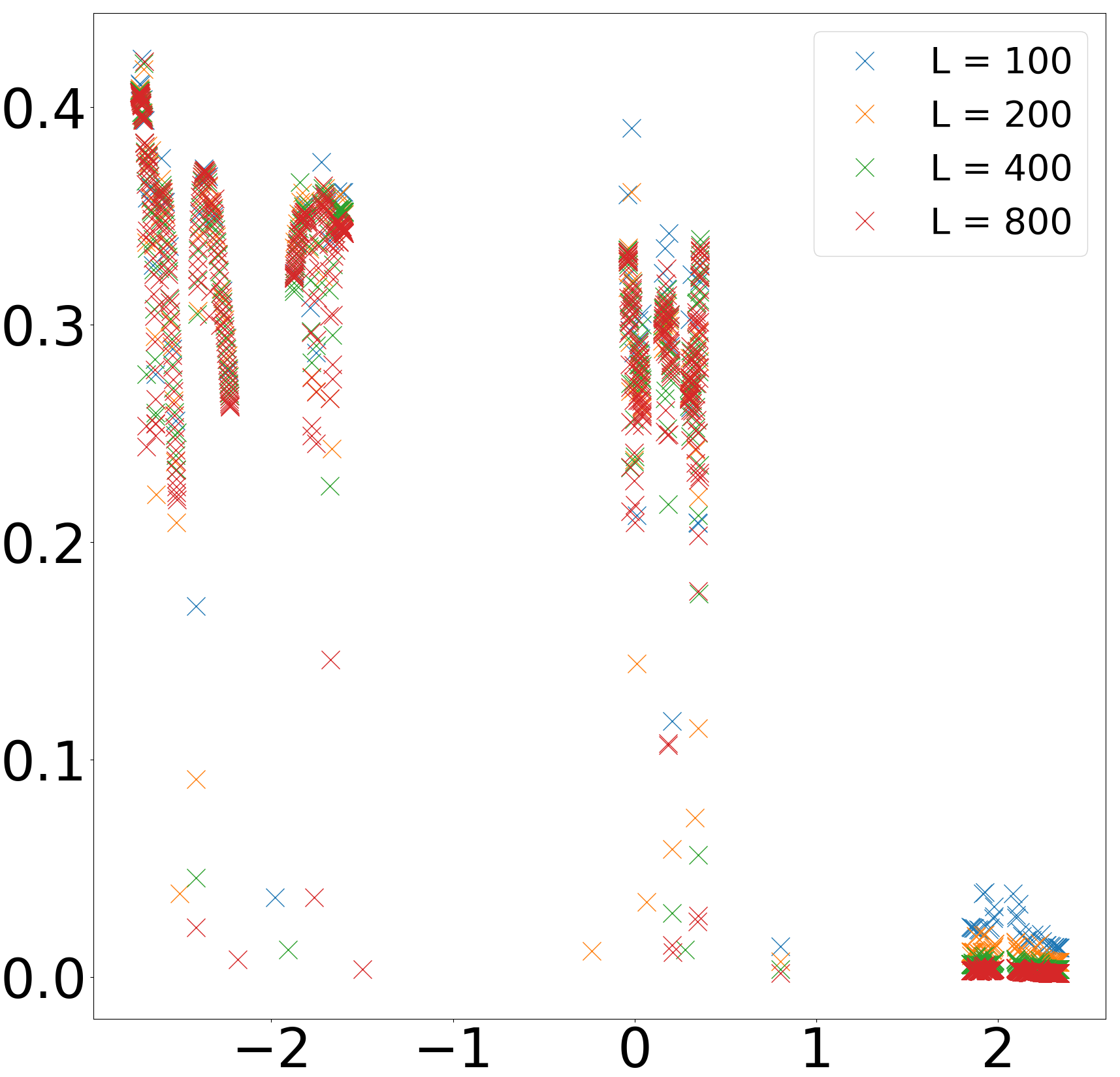}};
\node[left=of img23,node distance=0cm,rotate=90,anchor=center,yshift=-0.8cm]{\large{ $PR/L$}};
\node[left=of img23,node distance=0cm,yshift=2.5cm,xshift=3.8cm]{\large{b}};
\node[left=of img23,node distance=0cm,yshift=-2.5cm,xshift=4cm]{\large{$E$}};
\end{tikzpicture}
\end{minipage}
\begin{minipage}{0.3\textwidth}
\begin{tikzpicture}
\node (img24)[right= of img23,xshift = 12.2] {\includegraphics[width=0.9\textwidth]{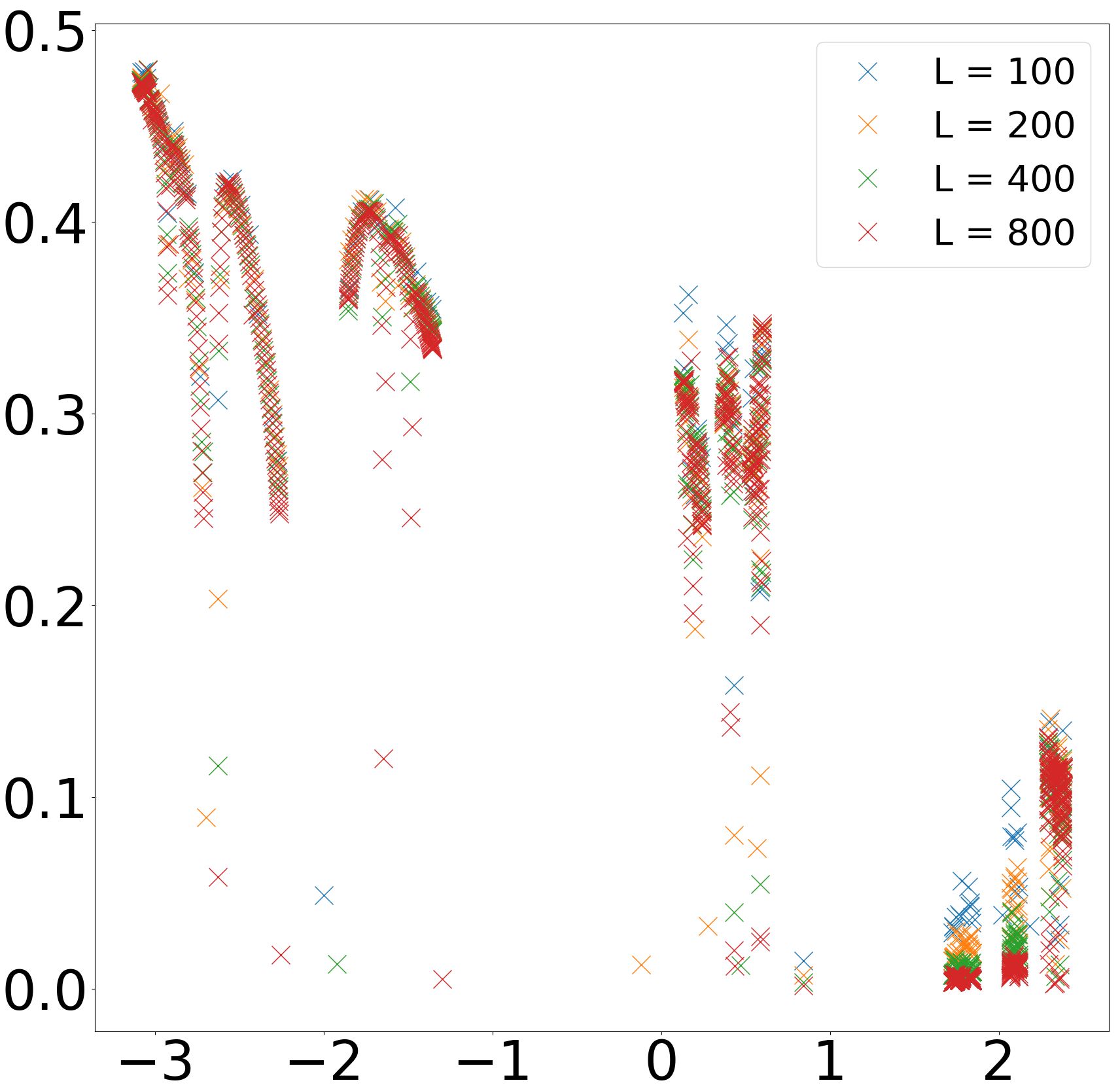}};
\node[left=of img24,node distance=0cm,rotate=90,anchor=center,yshift=-0.8cm]{\large{ $PR/L$}};
\node[left=of img24,node distance=0cm,yshift=2.5cm,xshift=3.8cm]{\large{c}};
\node[left=of img24,node distance=0cm,yshift=-2.5cm,xshift=4cm]{\large{$E$}};
\end{tikzpicture}
\end{minipage}
\begin{minipage}{0.3\textwidth}
\begin{tikzpicture}
\node (img25)[below= of img24, xshift = 10.2] {\includegraphics[width=0.9\textwidth]{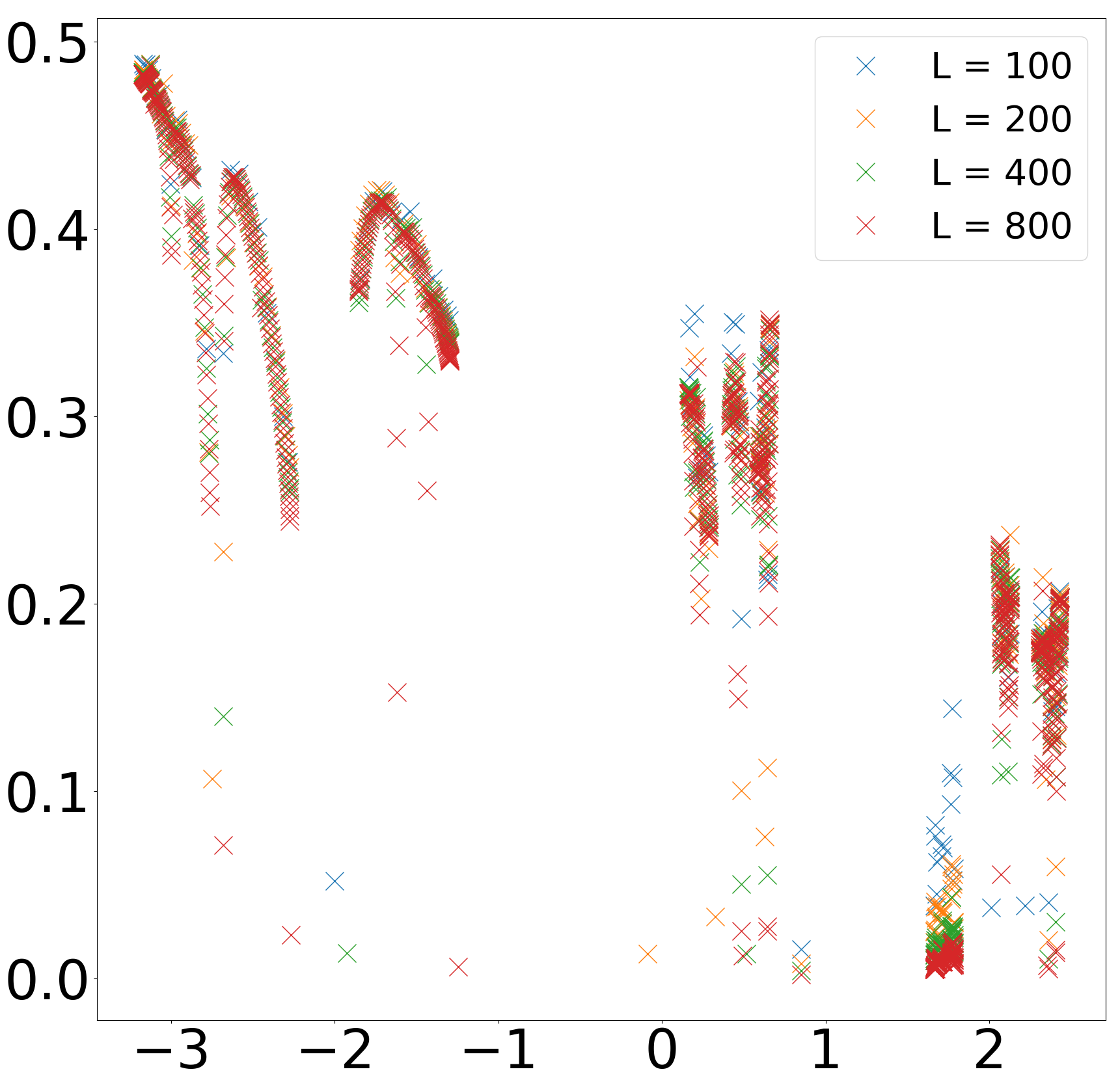}};
\node[left=of img25,node distance=0cm,rotate=90,anchor=center,yshift=-0.8cm]{\large{ $PR/L$}};
\node[left=of img25,node distance=0cm,yshift=2.5cm,xshift=3.8cm]{\large{d}};
\node[left=of img25,node distance=0cm,yshift=-2.5cm,xshift=4cm]{\large{$E$}};
\end{tikzpicture}
\end{minipage}
\begin{minipage}{0.3\textwidth}
\begin{tikzpicture}
\node (img26)[right= of img25,xshift = 10.2] {\includegraphics[width=0.9\textwidth]{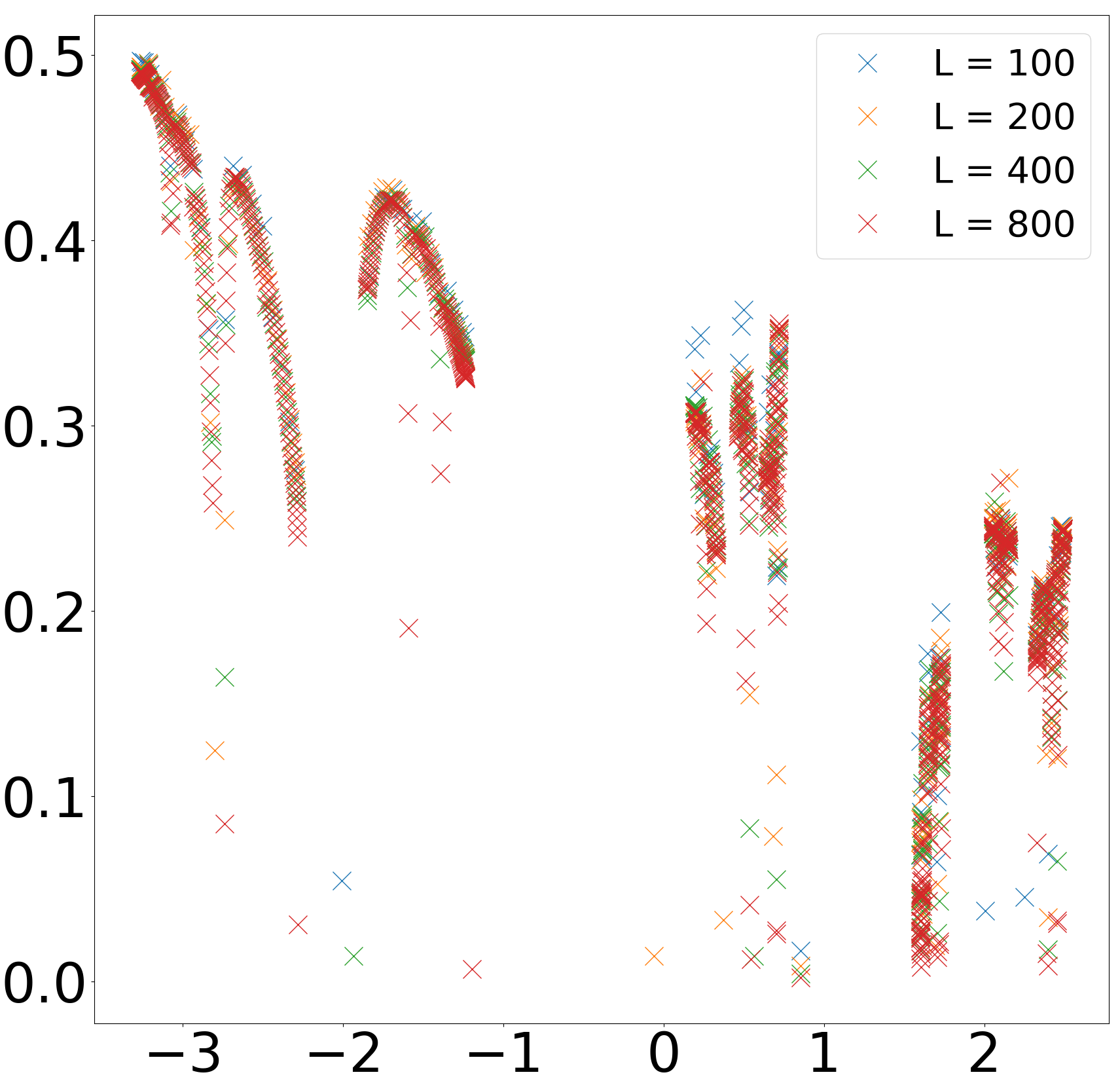}};
\node[left=of img26,node distance=0cm,rotate=90,anchor=center,yshift=-0.8cm]{\large{ $PR/L$}};
\node[left=of img26,node distance=0cm,yshift=2.5cm,xshift=3.8cm]{\large{e}};
\node[left=of img26,node distance=0cm,yshift=-2.5cm,xshift=4cm]{\large{$E$}};
\end{tikzpicture}
\end{minipage}
\begin{minipage}{0.3\textwidth}
\begin{tikzpicture}
\node (img27)[right= of img26,xshift = 10.2] {\includegraphics[width=0.9\textwidth]{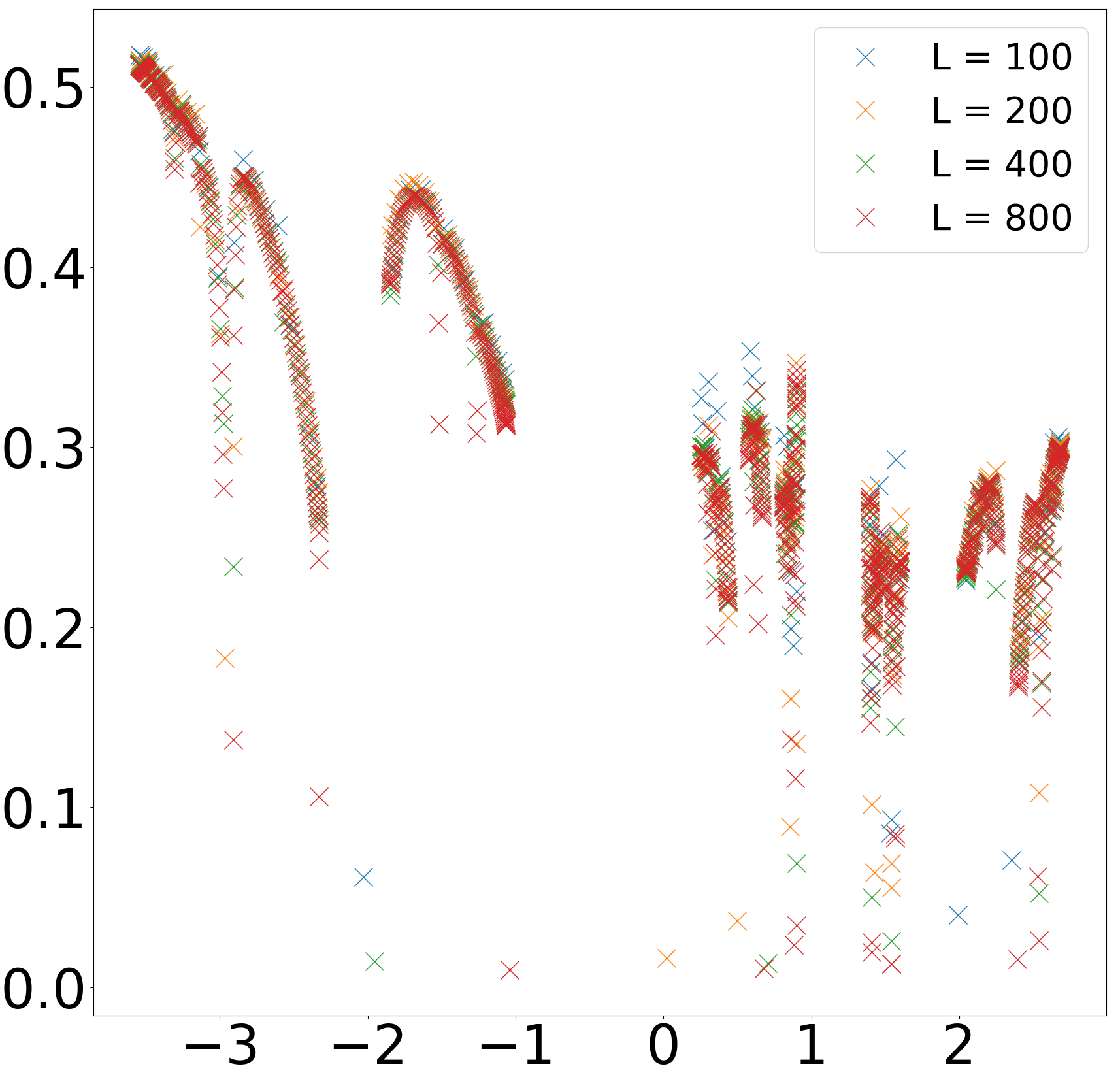}};
\node[left=of img27,node distance=0cm,rotate=90,anchor=center,yshift=-0.8cm]{\large{ $PR/L$}};
\node[left=of img27,node distance=0cm,yshift=2.5cm,xshift=3.8cm]{\large{f}};
\node[left=of img27,node distance=0cm,yshift=-2.5cm,xshift=4cm]{\large{$E
$}};
\end{tikzpicture}
\end{minipage}
\caption{ PR of single particle states, defined in Eq.~\ref{Part} for Model 3, scaled by system size. We recollect that PR is $O(1)$ for localized states and $O(L)$ for de-localized states (see Eq.~\ref{PRcases}). All the plots are for $\lambda = 0.9$, showing a transition from delocalized to mobility edge to delocalized phase again through appearances and merger of multiple mobility edges (a) $\alpha = 0$, the AAH limit, all states are delocalized. (b) $\alpha = 0.15$, a single mobility edge appears separating the $O(L)$ extended states from $O(1)$ localized states. For (c) $\alpha = 0.35$ and (d) $\alpha = 0.4$, two mobility edges appear separating a region of localized states in the spectrum around $E= 2$ from extended states on either sides. In other words, around $E=2$ there is a region of localized states. (e) $\alpha = 0.45$, the mobility edges merge to give  special band of states near $E = 1.6$  (f) $\alpha = 0.6$  the mobility edges have merged, giving a delocalized phase again.}
\label{IPR_appendix}
\end{figure*}
\vspace{0.5cm}

While Model 1 showed a non monotonicity of Imbalance in the localized phase and  mobility edge , this is not unique to Model 1, and can be  seen in the other models as well. Calculation of $I_{0}$ from the linear fits in Fig.~\ref{Finite_size_appendix} shows this non-monotonicity.
The phase diagram of Model 4, Fig.~\ref{Phase_diagram_M4} (a) shows a uniform localized phase for all values of $\alpha$ for $\lambda > 1$. However, the steady state imbalance showed that the nature of localized states are drastically different. They are very weakly localizing near the Bloch limit $\alpha = 0$, where they still retain some of their plane wave nature. This gives rise to a very low value of imbalance, and although localized, this model exhibits a value that is close to 0 even in the thermodynamic limit. The imbalance however spikes up close to $\alpha$ = 0.8, before dipping at $\alpha = 0.9$ before the model goes to the AAH limit . The nature for this sudden drop is not very clear yet.

\section{ Participation ratios for Model 3}\label{Mobilityedges_M3_appendix}

In Section \ref{Phases}, it was mentioned that the delocalization to mobility edge transition, of Model 3, and back to delocalized phase was due to appearance and merging of multiple mobility edges. These transitions may be seen from following any vertical cut in the region $0.75 < \lambda < 1.0$ in Fig.~\ref{Phase_diagram_M3} (b). The participation ratio, defined in Eq.~\ref{Part} describes the appearance  of mobility edges as an approximate energy value that separates the states with extensive and intensive scaling with system size, i.e., it is $O(L)$ for delocalized states, and $O(1)$ for localized states.   For various parameters this intricate transition is observed for Model 3, in the regime $\lambda < 1$. In Fig.~\ref{IPR_appendix}, one such transition is shown for $\lambda  = 0.9$. As we increase $\alpha$, the system moves from delocalized phase to a mobility edge phase. At higher $\alpha,$, a second mobility edge appears, and as $\alpha$ is increased further, these two edges merge together to transition the system back into the delocalized phase.

\bibliography{apssamp}

\end{document}